\documentclass[12pt,a4paper]{article}
\usepackage{jheppub_kim}
\usepackage{dirtytalk}
\usepackage{slashed}
\usepackage{longtable}
\usepackage{epsfig}
\usepackage{amssymb,amsmath}
\usepackage{amsthm}
\usepackage{graphicx}
\usepackage[hang,nooneline,scriptsize]{subfigure}
\usepackage{subfigure}
\usepackage{array}
\usepackage{color,soul}
\usepackage{wasysym}
\usepackage{caption}
\usepackage{float}
\usepackage{orcidlink}
\usepackage{comment}
\def \a{\alpha}

\begin{document}
  
\title{Unveilling Chaos in Particle Motion: Analyzing the Impact of Horizon in $f(R)$ Gravity}
\author[a]{Surajit Das,\orcidlink{0000-0003-2994-6951}}
\affiliation[a]{Physics and Applied Mathematics Unit, Indian Statistical Institute, 203, Barrackpore Trunk Road, Kolkata 700108, India.}
\author[b]{Surojit Dalui\orcidlink{0000-0003-1003-8451}\footnote{Corresponding author}}
\affiliation[b]{Department of Physics, Shanghai University, 99 Shangda Road, Baoshan District, Shanghai 200444, People's Republic of China.}

\emailAdd{surajit.cbpbu20@gmail.com} 
\emailAdd{surojitdalui@shu.edu.cn, surojitdalui003@gmail.com}

\abstract{This article is devoted to investigate the effects of $f(R)$ theory in the dynamics of a massless particle near the horizon of a static spherically symmetric (SSS) black hole. Deriving the equations of motion within $f(R)$ gravitational theories, novel solutions for charged and neutral black holes are obtained, introducing a dimensional parameter $a$ in $f(R)=R-2a\sqrt{R}$. Departing from General Relativity, these solutions showcase unique properties reliant on the dynamics of Ricci scalar. Analysis shows that chaos manifests within a specific energy range, with $a$ playing a crucial role. The study underscores the general applicability of the spherically symmetric metric, revealing insights into particle dynamics near black hole horizons. Despite an initially integrable nature, the introduction of harmonic perturbation leads to chaos, aligning with the Kolmogorov-Arnold-Moser theory. This research contributes to a nuanced understanding of black hole dynamics, emphasizing the importance of alternative theories of gravity. }

\keywords{Black hole; Event horizon; $f(R)$ gravity; Harmonic potential; Chaos.
}

\maketitle

\section{Introduction}\label{s1}
The understanding of the universe has been one of the prime objectives since the birth of science and with the progress of astronomy, astrophysics, cosmology, data science, and space science, the understanding has been more enriched by that. The latest observations strongly indicate that our universe is expanding at an accelerated rate \cite{Riess1,Perlmutter1,De,Colless,Perlmutter2,Spergel1,Peiris,Tegmark,Cole,Springel,Astier,Riess2,Spergel2,Ade,Komatsu}. This phenomenon is attributed to an unknown constituent known as `Dark Matter' (DM) and `Dark Energy' (DE). Numerous fundamental challenges, encompassing quantum gravity, dark energy, and dark matter, motivate the exploration of alternative gravitational theories beyond the conventional framework of Einstein's General Relativity (GR). General Relativity, while successful in many aspects, remains confronted with unresolved issues, such as singularities, the nature of DE and DM. These challenges prompt researchers to seek modifications or extensions of GR to address its limitations across both ultraviolet (UV) and infrared (IR) scales \cite{capozziello}. It is pertinent to consider the direct extension of GR, treating it as a special case within a broader gravitation theory. Among the various extensions, the $f(R)$ gravity theory represents an intriguing generalization of the Einstein-Hilbert action, replacing the Ricci scalar, $R$, with an analytic differentiable function. This approach is rooted in the requirements of formulating quantum field theories on curved spacetime \cite{birrell}. There is a wide range of successful applications of $f(R)$ gravity in the context of cosmology as well as from the astrophysical points of view. For general reviews on $f(R)$ gravity, see \cite{r1,r2,r3,r4} and references therein. Modified gravity theories need not necessarily recover GR in its entirety but can yield equivalent formulations, such as the teleparallel equivalent of general relativity. Conversely, extended gravity theories can exhibit GR as a limiting case, depending on specific choices or conditions \cite{cai}.

In recent times people have been closely examining a class of $f(R)$ gravity models (see, for instance, \cite{c1,c2,c3,c4,c5,c6,c7,c8} and the relevant literature). These models are a type of higher-order derivative gravity theory that deals with higher-order curvature invariants expressed as functions of the Ricci scalar, $R$.  The significance of these models lies in their ability to avoid Ostrogradski's instability \cite{Ostro}, a potential problem in general higher derivative theories \cite{Wood}. Constructing phenomenologically viable $f(R)$ gravity theories encounter several challenges, including instabilities within and beyond matter \cite{Dol,Sou}, vacuum stability \cite{Faraoni}, and constraints derived from established gravity properties in our solar system (refer to \cite{Chi,E,Clif} and related literatures). Additionally, figuring out the specific functional form of $f(R)$ from cosmological observations is also tricky because the background expansion doesn't uniquely determine it \cite{Mult1}. Several studies address these challenges by transforming the theory into a scalar-tensor framework and then examining solar system constraints through the parametrized post-Newtonian limit \cite{Dam,Mag}. In this light, one interesting starting point is to investigate a static spherically symmetric solution in the background of $f(R)$ gravitational theory to proceed further.

Therefore, in this context, it is worth mentioning that the exploration of black hole solutions within modified gravity frameworks, diverging from GR, assumes significance in discerning between distinct modified gravity theories and imposing constraints on model parameters through gravitational waves or shadows. The quest for precise solutions in the $f(R)$ theory of gravity is both crucial and challenging, given the intricate nature of the equations of motion with higher-order terms. Despite the complexity, numerous exact and numerical solutions have been successfully derived through diverse methodologies. Initiating with the simplest scenario, a specific class of $f(R)$ gravity featuring constant curvature has been examined. The solutions within this class, such as Schwarzschild-like \cite{mult1}, Reissner-Nordstr$\Ddot{o}$m-like \cite{dela,moon}, and Kerr-Newman-like solutions \cite{jar}, exhibit deviations from GR solutions only through a constant coefficient that can be assimilated into Newton’s constant. Additionally, static spherically symmetric solutions incorporating perfect fluid \cite{mult2}, Yang–Mills field \cite{sh1}, non-linear Yang–Mills field \cite{moon,sh2}, Maxwell field, and non-linear electromagnetic fields \cite{habib,holl,rod,hur} have been obtained. Employing Noether symmetries, axially symmetric solutions have been deduced from exact spherically symmetric solutions \cite{sal}. Furthermore, intriguing correspondences between solutions in Einstein conformally invariant Maxwell theory and those in $f(R)$ gravity devoid of matter fields in arbitrary dimensions are elucidated in \cite{hendi1,hendi2}. The investigation extends to spherically symmetric vacuum solutions in $f(R)$ gravity within higher dimensions, as detailed in \cite{amirabi,trp}.

Hence, in the context of $f(R)$ gravity, significant attention has been directed towards spherically symmetric black hole solutions, including the generator method \cite{amirabi} and the Lagrange multiplier method \cite{main1}. In this paper, the use of Lagrangian multipliers has facilitated the derivation of novel analytic solutions featuring a dynamic Ricci scalar and by employing the field equation of $f(R)$ gravity, it is obtained a new charged and neutral (charge-less) black hole solutions characterized by a dynamic Ricci scalar that asymptotically approaches flat spacetime \cite{main1,main2}. Notably, solutions involving dynamic curvatures \cite{mult1,main1} manifest distinctive characteristics compared to GR solutions. The motivation behind taking this approach is that it permits reducing the equations of motion to a single equation for a large class of metrics and that's why it is possible to construct an exact solution in the $f(R)$ theory of gravity. We can also recover all the known exact solutions by using a non-constant Ricci scalar by a non-trivial solution, reconstructed in $f(R)$ gravity \cite{saffari1,saffari2}. References \cite{cognola,vilenkin,capo} elucidate the application of the Lagrangian multipliers method for the systematic analysis of a Lagrangian system within the framework of the Friedmann-Lema$\hat{i}$tre-Robertson-Walker (FLRW) spacetime. Significantly, it is worth mentioning that this methodology stands out from conventional approaches of direct use of Einstein's field equations within the domain of $f(R)$ gravity.

On the other hand, when we talk about black hole solutions, the concept of event horizon comes into the picture automatically. This fundamental and enigmatic boundary of the black hole  represents a point beyond which classical events hold no sway over external observers. Black holes, theoretically, are the solutions derived from Einstein’s field equations, marking regions from which nothing can escape under classical principles. The recent detection of gravitational waves by LIGO has transformed black holes from a purely theoretical concept into real entities within our universe \cite{abbott1,abbott2,abbott3,abbott4}. The study of near-horizon physics, encompassing both classical and quantum aspects, holds significant importance. Within the classical framework, extensive research has explored the intriguing influence of the horizon on integrable systems, leading to their transition into chaotic states. This transformation has been successfully established in prior works \cite{i1,i2,i3,i4,i5,i6,i7,i8,i9,i10}. These studies encompass various scenarios, including spinning or magnetized black hole systems, and involve test particles of varying mass, charge, or spin. Recent quantum mechanical investigations have revealed that the presence of a killing horizon induces Brownian motion in particle trajectories when observed from an accelerated frame \cite{adhikari}. Additionally, a recent analysis \cite{hashimoto} investigates the impact of a Schwarzschild black hole on a massive test particle subjected to perturbations from harmonic potentials, accompanied by an extra potential to prevent the particle from crossing the horizon. In all instances, the motion of the particle exhibits chaotic characteristics. However, the situation for the chaotic nature of the particle trajectory in the context of modified gravity has not been addressed so far. Therefore, a pivotal inquiry in the domain of black hole physics emerges: \textit{Is it possible to determine the chaotic dynamics of particle motion as it approaches the event horizon within the framework of modified theories of gravity?} Given that modified gravity theories present an alternative and well-established gravitational paradigm distinct from Einstein's GR. Therefore investigating the impact of chaotic trajectories of particles by event horizon in a specific modified theory of gravity holds significant interest. Hence the study of the physics near the event horizon emerges as a crucial avenue, capable of shedding light on numerous mysteries pertaining to the properties and behaviors of black holes in the context of modified theories of gravity.

In this paper, our focus is on comprehending the behavior of particles as they approach the event horizon of a black hole in a particular $f(R)$ gravity, with an exclusive emphasis on classical analysis. This paper offers a comprehensive exploration of a fundamental issue in particle dynamics near black hole horizons in $f(R)$ gravity. Our calculations reveal that when a particle approaches the black hole event horizon, its outgoing radial trajectory exhibits exponential growth over time. This observation suggests the possibility of inducing chaotic behavior in the motion dynamics of particles, particularly when the particle is initially part of an integrable system. We conducted a thorough numerical analysis of the Poincaré sections depicting particle trajectories influenced by harmonic potentials and the black hole event horizon. We considered two types of scenarios: one involving static spherically symmetric (SSS) charged black holes and another featuring SSS charge-less or neutral black holes in the context of $f(R)$ background. Our findings indicate that a particle confined by a harmonic potential can be categorized as an integrable system, but its nature of trajectory undergoes a significant transformation as it approaches the black hole horizon. In particular, we observed that both charged and neutral static spherically symmetric horizon induces chaos within a specific energy range, and the presence of a dimensional parameter (which is identified due to incorporating of reconstruction of $f(R)$ gravity) intensifies the system’s chaotic behavior.

We now turn our attention to the implications of our current analysis. Our study focuses exclusively on massless particles following outgoing trajectories, specifically the radial null geodesic. Notably, this same null geodesic plays a pivotal role in Hawking radiation as a tunneling effect \cite{parikh,rabin} radiating from the black hole’s event horizon. Remarkably, our findings suggest that particles emitted from the horizon, once liberated from its gravitational grasp, exhibit chaotic motion. This chaotic behavior is attributed to the dual influence of the event horizon and external perturbations introduced by various celestial objects in the Universe. This observation hints at the intriguing idea that the black hole horizon, in addition to emitting Hawking radiation, also imparts chaotic dynamics to the radiated particles. Simultaneously, it is intriguing to observe that the existence of a black hole's horizon induces chaotic trajectories in particle motion within the framework of modified gravity, specifically in the context of $f(R)$ gravity. This observation suggests that the manifestation of chaos as a consequence of the presence of a horizon is a universal phenomenon under the consideration of $f(R)$ gravity. In our discourse, it is important to clarify that we do not treat the massless particle as an inherent harmonic oscillator; instead, we consider a scenario in which the particle is confined within a harmonic potential. This perspective is not novel and finds precedent in earlier studies, including the examination of phenomena such as the propagation of optical beams subjected to harmonic potentials \cite{zhang} and the analysis of relativistic massless harmonic oscillators \cite{kow}.

This work is presented as follows. In Section \ref{s2a}, we give a brief overview of the Lagrangian approach for the derivation of equations of motion in $f(R)$ gravity, which is followed by two subsections. In the Subsection \ref{s2b}, we study the solution for the charged black hole in the reconstructed $f(R)$ gravity background and in the Subsection \ref{s2c}, the charge-less black hole solution has been derived. Within Section \ref{sec3}, we have analysed the trajectories of massless particle in a very near horizon region and also discussed the dynamical equations of motion, derived from the dispersion relation. As a whole in Section \ref{sec4}, we have studied and numerically observed the various Poincar$\Acute{e}$ sections of the particle trajectories for the charged and charge-less black holes. Finally, we summarize our results and final remarks in Section \ref{sec5}.

\section{Background as a reconstruction of $f(R)$ gravity}\label{sec2}
In this section, we will discuss a static spherically symmetric solution in $f(R)$ gravity as a background spacetime. We'll investigate two types of  static spherically symmetric solutions in $f(R)$ gravity. One is a charge-less or neutral black hole solution and another is the charged case. But before proceeding further we need to know one more i.e. the choice of the generic function $f(R)$ in the background spacetime. Therefore in the next subsection \ref{s2a} we've discussed the SSS solution in $f(R)$ gravity from the Lagrangian approach and then in subsection \ref{s2b}, we'll discuss the solution of charged black hole for a chosen $f(R)$ given in the subsection \ref{s2a}. Finally, the neutral black hole solution is discussed in the subsection \ref{s2c}.

\subsection{Lagrangian derivation of equation of motion}\label{s2a}
The generic action for $f(R)$ gravity is given by,
\begin{equation}
    \mathcal{A}_{g}=\int d^4x\sqrt{-g}f(R),\label{2.1}
\end{equation}
where $g$ is the determinant of the fundamental metric tensor $g^{\mu\nu}$ in a $4D$ spacetime. Now a general static spherically symmetric metric in a $4D$ spacetime can be written as,
\begin{equation}
    dS^2=-P(r) e^{2\a(r)}dt^2+\frac{dr^2}{P(r)}+r^2d\Omega^2.\label{2.2}
\end{equation}
Here $\a(r)$,~$P(r)$ are interesting to calculate which are an unknown functions of $r$ and $d\Omega^2=d\theta^2+\sin^2\theta d\phi^2$.

In the references \cite{cognola,vilenkin,capo}, it has been shown that we can study the employment of the  Lagrangian multipliers method to effectively analyze a Lagrangian system in the paradigm of the  Friedmann-Lemaitre-Robertson-Walker (FLRW) spacetime, where the Lagrangian involves first order derivatives of generalized coordinates with the corresponding generalized velocities as well. L. Sebastiani et. al. have studied this particular Lagrangian methodology which facilitates the consideration of scalar curvature $R$, along with the relevant quantities, $\a(r)$ and $P(r)$, associated with the static spherically symmetric ansatz \eqref{2.2}, as independent Lagrangian coordinates \cite{main1}. As a results, this approach yields two distinct equations of motion, with one of the unknown variables, $\a$, featuring in a straightforward manner. Notably, this approach distinguishes itself from other conventional techniques by its deliberate avoidance of a direct reliance on the Einstein's field equation in the context of $f(R)$ gravity.

Now the action \eqref{2.1} by introduced the Lagrangian multipliers $\lambda$ can be written as follows \cite{safko,main1}:
\begin{eqnarray}
    \mathcal{A}_{g}\equiv\frac{1}{2\kappa^2}\int dt\int e^{\a(r)}r^2 dr \Bigg[f(R)-\lambda\Bigg(R+\Big(3P'(r)\a'(r)+2P(r){\a^\prime}^2(r)+P''(r)\nonumber\\
    +2P(r)\a''(r)+4\frac{P'(r)}{r}+4\frac{P(r)\a'(r)}{r}+2\frac{P(r)}{r^2}-\frac{2}{r^2}\Big)\Bigg)\Bigg],\label{2.3}
\end{eqnarray}
where we've used the value of the curvature scalar using \eqref{2.2}, given by
\begin{eqnarray}
    R=&&-3P'(r)\a'(r)-2P(r){\a^\prime}^2(r)-P''(r)-2P(r)\a''(r)\nonumber\\
    &&-4\frac{P'(r)}{r}-4\frac{P(r)\a'(r)}{r}-2\frac{P(r)}{r^2}+\frac{2}{r^2}~.\label{2.4}
\end{eqnarray}
Here $\prime$ denotes the derivative with respect to the radial coordinate $r$. In this study, one point must be noted that we introduce an approach for the exploration of SSS solutions within the framework of a generic $f(R)$ theory. We posit that this method demonstrates notable generality and offers distinct advantages over other established techniques, given in the references \cite{mult1,mult2,saffari1,saffari2,capo1,trp}. Notably, in the significant scenario of a constant $\a(r)$ (where we'll set $\a$ to zero in our further studies), we derive a comprehensive metric form \eqref{2.2} by means of an explicit expression for the quantity denoted as $P(r)$. Moreover, this specific metric form serves as a compelling impetus for the pursuit of a comprehensive reconstruction of $f(R)$ gravity that encompasses the differentiations between standard GR and the modifications of GR. The inclusion of this parameter potentially offers a novel avenue for elucidating a broader spectrum of ramifications associated with the examination of black hole horizons in the context of $f(R)$ geometry.\\
Now, for knowing the value of undetermined multiplier $\lambda$, we make the variation of the given action \eqref{2.3} with respect to $R$, one gets,
\begin{eqnarray}
    \frac{df(R)}{dR}=\lambda\label{2.5}.
\end{eqnarray}

Finally by making an integration by parts and using \eqref{2.5}, one can obtain the Lagrangian of the system in this context, which takes the form as follows:
\begin{eqnarray}
    \mathcal{L}=e^{\a(r)}\Bigg[r^2\Big(f(R)-Rf_R(R)\Big)+2f_R(R)\Big(1-P(r)-rP'(r)\Big)\nonumber\\
    +f_{RR}(R)r^2R'(r)\Big(P'(r)+2P(r)\a'(r)\Big)\Bigg],\label{2.6}
\end{eqnarray}
where $f_R(R),~f_{RR}(R)$ denote the first and second order derivatives of the arbitrary analytic function $f$ with respect to $R$, respectively. It is noteworthy to mention that the generic Lagrangian, given in \eqref{2.6}, in our case is basically a function of three variables i.e., $\mathcal{L}\equiv\mathcal{L}\big(\a(r),\a'(r),R(r),R'(r),P(r),P'(r)\big)$. Therefore we can get three equations of motion for making the variation of three dynamical variables such as $\a(r),R(r)$ and $P(r)$, respectively.\\
Now for the first and third equations of motion, taking the variations of the Lagrangian \eqref{2.6} with respect to $\a(r)$ and $P(r)$, respectively, we get the following:
\begin{eqnarray}
    &&\frac{d^2R}{dr^2}+\Big[\frac{2}{r}+\frac{P'(r)}{2P(r)}\Big]\frac{dR}{dr}+\frac{f_{RRR}}{f_{RR}}{\Big(\frac{dR}{dr}\Big)}^2+\frac{Rf_R-f(R)}{2P(r)f_{RR}}\nonumber\\
    &&-\frac{f_R}{r^2 P(r)f_{RR}}\Big[1-P(r)-rP'(r)\Big]=0~,\label{2.7}\\
    &&{f_{RR}}\frac{d^2 R}{dr^2}-\a'(r)\Big(\frac{2f_R}{r}+f_{RR}{\frac{dR}{dr}}\Big)+f_{RRR}{\Big(\frac{dR}{dr}\Big)}^2=0.\label{2.8}
\end{eqnarray}
We'll use these equation further for our studies. We must mention that the Lagrangian approach in this context is correct as we can recover the Ricci curvature scalar \eqref{2.4} by taking the variation of \eqref{2.6} with respect to $R$ as a second equation of motion. An one strong motivation behind taking the Lagrangian approach unlike taking standard GR approach is that the variable $P(r)$ does not appear in the third equation of motion \eqref{2.8} and also for the first equation of motion \eqref{2.7} which consist the $P(r),P'(r)$ along with $f(R)$ and their higher derivative terms will not contain any terms related to $\a(r)$. So we may say that, by taking this particular approach one can reduce the equation of motion. 

For considering non-constant Ricci curvature scalar and constant $\a$, we have the following solution of \eqref{2.8} as,
\begin{equation}
    f_R(R)=\gamma r+\delta,\label{2.9}
\end{equation}
where $\gamma$ and $\delta$ are the arbitrary integration constants. For the realization of the reconstruction of $f(R)$ gravity model, we have to express the Ricci scalar $R$ as a function of $r$ so that we may have the explicit form of $R$ from \eqref{2.4} with constant $\a$.
\begin{equation}
    \frac{d^2P(r)}{dr^2}+\frac{4}{r}\frac{dP(r)}{dr}+2\frac{P(r)}{r^2}-\frac{2}{r^2}=-R.\label{2.10}
\end{equation}
To find out the functional form of $P(r)$ as followed by the article \cite{main1}, from \eqref{2.7} with using \eqref{2.9}, \eqref{2.10} and the identities  $\frac{df_R}{dr}=\gamma\equiv f_{RR}\frac{dR}{dr}$, we can have
\begin{equation}
    \Big(\gamma+\frac{\delta}{r}\Big)\frac{d^2P(r)}{dr^2}-\frac{2\gamma}{r^2}\Big(2P(r)-1\Big)-\frac{2\delta}{r^3}\Big(P(r)-1\Big)+\frac{\gamma}{r}\frac{dP(r)}{dr}=0.\label{2.11}
\end{equation}
As $\delta$ is adimensional, choosing $\delta=1$, the general solution of \eqref{2.11} having the form,
\begin{eqnarray}
    P(r)=\frac{1}{2}\Big[2+r\big(2rC_1+3\gamma^2 r-2\gamma\big)+\gamma C_2\big(1-2\gamma r\big)\Big]\nonumber\\
    -\gamma^2 r^2\Big[\ln r-\ln(1+\gamma r)\Big]\big(1+\gamma C_2\big)-\frac{C_2}{3r}~,\label{2.12}
\end{eqnarray}
where $C_1$ and $C_2$ are two generic integration constants. So now for taking $\a=0$, if $f(R)$ gravity model realizes the metric 
\begin{equation}
    dS^2=-P(r)dt^2+\frac{dr^2}{P(r)}+r^2d\Omega^2,\label{2.13}
\end{equation}
then the coefficient $P(r)$ must satisfies the generic form of \eqref{2.12} as a general form of $P(r)$ \cite{main2}. Here the beauty of the metric ansatz \eqref{2.2} lies as a more general form for a reconstruction of modified gravity as $f(R)$ background.

Now it could be chosen for the generic constants $C_1=-\frac{3\gamma^2}{2}$, and $C_2=-\frac{1}{\gamma}$, we have the general solution for $P(r)$ as \cite{main1},
\begin{equation}
    P(r)=\frac{1}{2}+\frac{1}{3\gamma r},\label{2.14}
\end{equation}
and the Ricci scalar reads as inverse square of $r$ from \eqref{2.10}. Finally, by taking the above chosen constants $C_1$ and $C_2$, we have the general solution from \eqref{2.9} as,
\begin{equation}
    f(R)=R+2\gamma\sqrt{R}\label{2.14a}
\end{equation}
One point must be noted that from \eqref{2.14} we have $r_H=-\frac{2}{3\gamma}$, but for the physical solution of black hole horizon radius, the parameter $\gamma$ should be positive and must be greater than zero.  As a result, in this context, the reconstruction of $f(R)$ gravity model is given by \cite{main1}
\begin{equation}
    f(R)=R-2a\sqrt{R} ,~~~~\gamma=-a<0 ;~~a>0.\label{2.15}
\end{equation}

\subsection{Charged black hole solution}\label{s2b}
To get the charged black hole solution, we have to couple electromagnetic field with gravitational action in the background of $f(R)$ gravity, which follows the action,
\begin{eqnarray}\label{2.16}
    \mathcal{A}_{total}=\mathcal{A}_{g}+\mathcal{A_{EM}}=\int d^4x\sqrt{-g}\Big[\frac{1}{2\kappa}f(R)-\frac{1}{2}F_{\mu\nu}F^{\mu\nu}\Big],
\end{eqnarray}
where $F_{\mu\nu}$ is the standard electromagnetic field tensor, defined as $F_{\mu\nu}=\partial_{\mu}A_{\nu}-\partial_{\nu}A_{\mu}$ with $A_{\mu}$ is the one-form gauge potential. By varying the corresponding action with respect to the metric tensor $g_{\mu\nu}$, we obtain the field equations of $f(R)$ gravitational theory.
\begin{eqnarray}
    &&I_{\mu\nu}\equiv R_{\mu\nu}f_{R}-\frac{1}{2}f(R)g_{\mu\nu}+\big(g_{\mu\nu}\square-\nabla_{\mu}\nabla_{\nu}\big)f_R-\kappa T_{\mu\nu}=0,\label{2.17}\\
    && \text{with}~,\nonumber\\
    &&\frac{1}{\sqrt{-g}}\partial_{\mu}(\sqrt{-g}F^{\mu\nu})=0.\label{2.18}
\end{eqnarray}
The trace of the field equations \eqref{2.17} is given by,
\begin{eqnarray}
    Rf_R-2f(R)+3\square f_R=\kappa T\label{2.19}.
\end{eqnarray}
Here it should be mentioned that the Maxwell field is conformally invariant which means the property of vanishing of the trace of Maxwell field energy-momentum tensor $T$ and in this context, the energy-momentum tensor of the Maxwell field is defined as,
\begin{eqnarray}
    T_{\mu\nu}\equiv\frac{1}{4\pi G}\Big(g_{\sigma\epsilon}F^{\sigma}_{\nu}F^{\epsilon}_{\mu}-\frac{1}{4}g_{\mu\nu}F_{\sigma\epsilon}F^{\sigma\epsilon}\Big).\label{2.20}
\end{eqnarray}

\begin{figure}[ht]
\begin{center}
\includegraphics[width=0.8\linewidth]{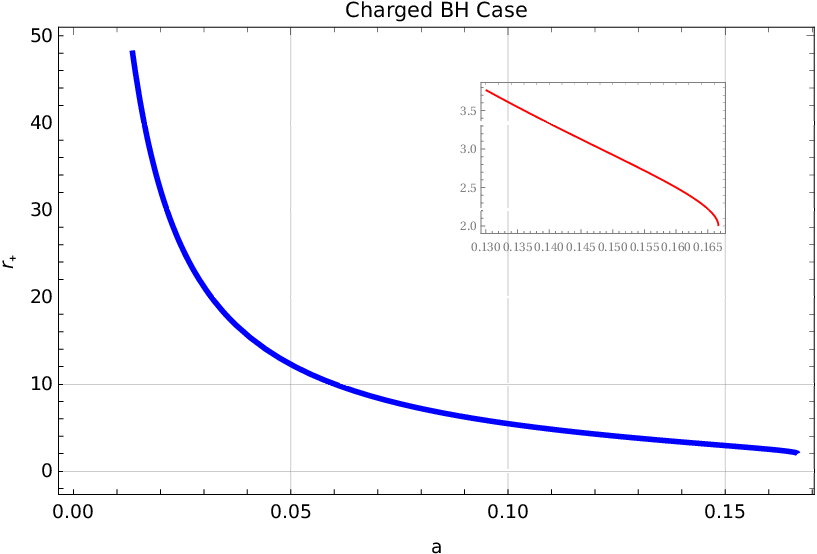}
\end{center}
\caption{Plot of dimensional parameter $a$ versus the outer horizon radius $r_+$}
\label{f1}
\end{figure}

Now let us derive a charged black hole solution considering the choice of the $f(R)$ gravity model as given in \eqref{2.15}. So assuming the SSS background spacetime as given in \eqref{2.13} and using the field equations \eqref{2.17}, \eqref{2.18}, and \eqref{2.19} by solving the system of equations $(I^t_t-I^r_r)$ and $I^{\theta}_{\theta}$ and also using \eqref{2.10}, we get the following charged black hole solution for the metric \eqref{2.13} as:
\begin{eqnarray}
    dS^2=-\Big(\frac{1}{2}-\frac{1}{3ar}+\frac{1}{3ar^2}\Big)dt^2+{\Big(\frac{1}{2}-\frac{1}{3ar}+\frac{1}{3ar^2}\Big)}^{-1}dr^2+r^2 d\Omega^2.\label{2.21}
\end{eqnarray}
It is worthy to mention here that the Ricci scalar satisfies the inverse square law i.e., $R=\frac{1}{r^2}$ as we've adopted the $f(R)$ gravity model as $f(R)=R-2a\sqrt{R}$. The above metric solution behaves like asymptotically flat spacetime and this particular solution is similar with the procedures and results obtained by Nashed et. al. in their Ref. \cite{main2}.\\
For the physical solution of the charged black hole horizon by solving the equation $P(r)\equiv\Big(\frac{1}{2}-\frac{1}{3ar}+\frac{1}{3ar^2}\Big)=0$, we get the horizon radius as,
\begin{eqnarray}
    r_{\pm}=\frac{1\pm\sqrt{1-6a}}{3a},\label{2.22}
\end{eqnarray}
where \lq$+$\rq denotes the outer horizon and \lq$-$\rq denotes the inner horizon radius respectively. The choice of the dimensional parameter $a$ must be unique since for $a>0.166$, $r_{\pm}$ will give imaginary. Again the dimensional parameter $a$ must be greater than zero due to the validations of the equations \eqref{2.17}, \eqref{2.18}, and \eqref{2.19}. Hence in this case, the parameter $a$ should be in the range as $0<a\leq 0.166$. The graphical representation of $a$ vs. the outer horizon radius is shown in Fig.\ref{f1}. In this paper, we'll use \eqref{2.22} with outer horizon radius as a charged black hole solution for our further study.

\subsection{Neutral black hole solution}\label{s2c}
\begin{figure}[ht]
\begin{center}
\includegraphics[width=0.8\linewidth]{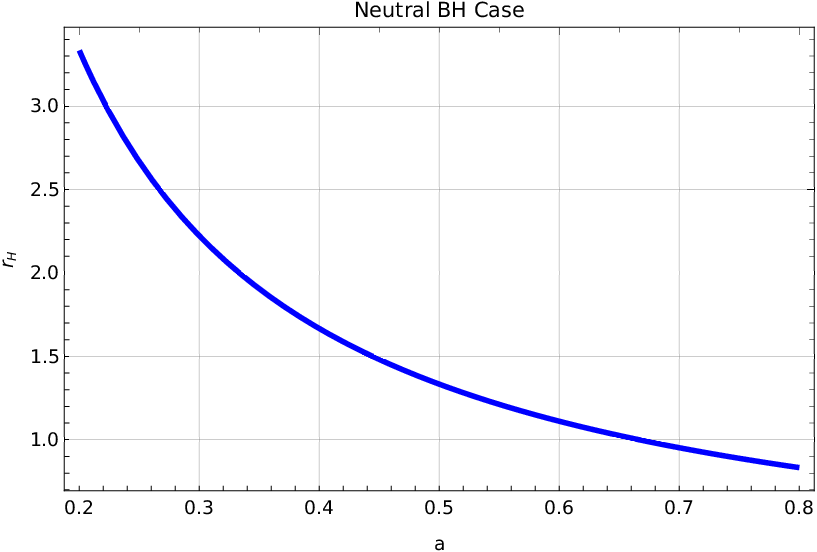}
\end{center}
\caption{Plot of dimensional parameter $a$ versus the horizon radius $r_H$ for the neutral black hole}
\label{f2}
\end{figure}

It is easy to find the charge-less or neutral black hole solution in the considered $f(R)$ gravity model as $f(R)=R-2a\sqrt{R}$, where the dimensional parameter $a$ must be positive. The exact solution for the neutral black hole adopting the SSS ansatz \eqref{2.13} is given by
\begin{eqnarray}
    P(r)=\frac{1}{2}-\frac{1}{3ar},\label{2.23}
\end{eqnarray}
which is already derived in the subsection \ref{s2a}.\\
In the case of charged black hole solution, the term $\frac{1}{3ar^2}$ \big(given in \eqref{2.21}\big) is responsible for the electric charge. Therefore neglecting the term $\frac{1}{3ar^2}$ will give us the charge-less black hole solution in $f(R)$ gravity, which is consistent with \eqref{2.14}. We'll use this equation \eqref{2.23} as a charge-less black hole solution for our study.\\
By solving $P(r)\equiv\Big(\frac{1}{2}-\frac{1}{3ar}\Big)=0$, we get the neutral black hole horizon radius $r_H=\frac{2}{3a}$. Here $a$ must be greater than zero and doesn't have any bounded region but one thing must be observed that for a very large values of $a$, the size of the black hole will be small. The graph of $a$ with horizon radius $r_H$ is plotted in Fig.\ref{f2}. 

\section{Equations of motion in SSS black hole geometry}\label{sec3}
In this section, we will discuss briefly on the various forms of equations of motion for a static spherically symmetric (SSS) black hole geometry. Considering a SSS black hole background as,
\begin{equation}\label{3.1}
    ds^2=-A(r)dt^2+\frac{dr^2}{A(r)}+r^2 d\Omega^2,
\end{equation}
where $d\Omega^2=d\theta^2+\sin^2\theta d\phi^2$ is the line element of a $2D$ unit sphere. Now it is easily seen that this black hole has a event horizon singularity at $r=r_H$ by the solution of $A(r)=0$. Since the metric does not depend explicitly on $t$, to remove the coordinate singularity at this position, we consider a better choice of coordinate system i.e., Painlev$\Acute{e}$-Gullstrand coordinates \cite{blau} by,
\begin{equation}\label{3.2}
    dt\longrightarrow dt-\frac{\sqrt{1-A(r)}}{A(r)} dr.
\end{equation}
Using \eqref{3.2}, the metric \eqref{3.1} takes the following form as,
\begin{equation}\label{3.3}
    ds^2=-A(r)dt^2+2\sqrt{1-A(r)}dt dr+dr^2+r^2 d\Omega^2.
\end{equation}
Here, it is important to mention that this spacetime has a time-like Killing vector $\zeta^{\alpha}=(1,0,0,0)$, so that it's energy is given by $E=-\zeta^{\alpha}p_{\alpha}$, where $p_{\alpha}$ is the four momentum vector under this background. Next, to find the energy of a particle in terms of other components of four momentum, we use the relativistic Hamilton-Jacobi equation which is nothing but the covariant form of the dispersion relation given by,
\begin{equation}\label{3.4}
    g^{\alpha\beta}p_{\alpha}p_{\beta}=-m^2 c^2,
\end{equation}
with m, mass of the particle.

Now using the dispersion relation \eqref{3.4} (with $c=1$) for the metric \eqref{3.3}, we have the energy of a massless particle as the following.
\begin{equation}\label{3.5}
    E=-\sqrt{1-A(r)}p_r\pm\sqrt{p^2_{r}+\frac{p^2_{\theta}}{r^2}}.
\end{equation}
Here, we've assumed the particle is moving only along the radial ($r$) and the $\theta$ directions only. The negative sign denotes the energy for the ingoing particle, while, the positive sign is for the outgoing particle. As the event horizon of a black hole behaves like a one-way membrane, in this paper,  throughout the discussions, we are mainly interested to study it's dynamics for the outgoing particle only. 

Now a general question arise that \textit{what will happen for the moment when the particle has a very near horizon radial motion?} To answer it, let us now take the equation for the energy \eqref{3.5} with the choice  of $p_{\theta}=0$, which is given by,
\begin{equation}\label{3.6}
    \dot{r}=\frac{\partial E}{\partial p_r}=1-\sqrt{1-A(r)}\simeq\kappa(r-r_H),
\end{equation}
where we've considered only the first-order expansion of $A(r)$ near the horizon as $A(r)\simeq 2\kappa(r-r_H)$. The term $\kappa=\frac{A'(r_H)}{2}$ is the surface gravity of the black hole.

It should be mentioned that in the context of Hawking radiation as a tunneling \cite{parikh,rabin}, the radial null geodesic i.e. the above equation has a very interesting connection with the Hawking effect. This path has been used to find the tunneling probability from the event horizon for the outgoing particles and one finds that this is non-zero, which leads to the Hawking radiation as a temperature, given by $T_H=\frac{\hbar\kappa}{2\pi}$. So one can say that for the radiated particles after escaping from the event horizon, it may exhibit some interesting characteristics in its motion due to the impact of the event horizon as well as the perturbation induced by the presence of other objects in the universe. 

Now the solution of the radial geodesic \eqref{3.6} is given by,
\begin{equation}\label{3.7}
    r=r_H+\Bar{c} r_H e^{\kappa\tau},
\end{equation}
where $\Bar{c}$ is the integration constant and $\tau$ is the affine parameter. On the other hand, considering the leading order equation of the radial momentum in the near-horizon region,
\begin{equation}\label{3.8}
    \dot{p_r}\simeq-\kappa p_r,
\end{equation}
which leads to the solution of $p_r$ as $p_r=p_{r_0}e^{-\kappa\tau}$ ($p_{r_0}$ is an arbitrary constant). Hence it is clear from the solutions of the equations \eqref{3.7}, \eqref{3.8} that either $r$ or $p_r$ shows the exponential growth with the increase of the affine parameter $|\tau|$ and here we consider $\tau$ to be positive. We can say that a photon that can have unstable circular motion in general relativity, but not expected in Newtonian approximation \cite{carroll}. Therefore under the influence of the horizon, the exponential growth of the radial motion can be signified as the presence of chaos in an integrable system. The value of the maximum Lyapunov exponent in this context is defined as \cite{maldacena},
\begin{equation}\label{3.9}
    \tau_{L,max}=\lim_{\tau\to\infty}\frac{1}{\tau}\ln\Big(\frac{\delta r(\tau)}{\delta r(0)}\Big).
\end{equation}
The term $\delta r(\tau)$ represents the separation between two infinitesimal close trajectories at time parameter $\tau$. Since we are considering the motion of null geodesics, so that the proper time can't be used as a valid parameter as it vanishes along the null path. So in this paper, we've chosen the four momentum $p^\mu$ in such a way that \eqref{3.5} is well defined and the parameter $\tau$ turns out to be affine one. The affine parameter $\tau$ should be a scalar and invariant under coordinate transformation. This implies that the Lyapunov exponent
defined here does not has any coordinate dependence. In this situation, the Lyapunov coefficient $\tau_{L}$ is bounded as,
\begin{equation}\label{3.10}
    \tau_{L}\leq\frac{2\pi T_H}{\hbar},
\end{equation}
which was first mentioned for the Sachdev–Ye–Kitaev (SYK) model by the authors Maldacena et al. \cite{maldacena}.

Let us now briefly discuss on the equations of motion and the actual trajectories of the particle. Here we want to mention that in a four-dimensional phase space the actual trajectories of the particle are highly non-linear in nature. In this paper, our motivation is to study the collective effect of the black hole event horizon on the nature of particle motion or the particle trajectories. So we are assuming a situation where particle is trapped under two harmonic potentials $\frac{1}{2}K_r(r-r_c)^2$ and $\frac{1}{2}K_\theta(y-y_c)^2$ along $r$ and $\theta$ directions, respectively. The terms $K_r$, $K_\theta$ represent the spring constants along $r$ and $\theta$ directions, respectively, where $y=r_{H}\theta$ and $r_c$ and $y_c$ are the equilibrium positions of these two harmonic potentials. Such model was suggested in these references \big(\cite{hashimoto} for massive particle and \cite{dalui,Dalui:2019umw} for massless particle\big). Therefore, the massless particle is initially under these type of harmonic potentials and thus the whole system is kept under the influence of the black hole horizon as well.

Now the total energy of this particle under the influence of harmonic potentials for the metric \eqref{3.3} is given by,
\begin{equation}\label{3.11}
    E=-\sqrt{1-A(r)}p_r+\sqrt{p^2_{r}+\frac{p^2_{\theta}}{r^2}}+\frac{1}{2}K_r(r-r_c)^2+\frac{1}{2}K_\theta(y-y_c)^2.
\end{equation}
Correspondingly, the equations of motion have the following forms:
\begin{eqnarray}
    &&\dot{r}=\frac{\partial E}{\partial p_r}=-\sqrt{1-A(r)}+\frac{p_r}{\sqrt{p^2_r+\frac{p^2_{\theta}}{r^2}}},\label{3.12}
    \\
    &&\dot{p_r}=-\frac{\partial E}{\partial r}=-\dfrac{A'(r)}{2\sqrt{1-A(r)}}p_r+\dfrac{p^2_{\theta}/r^3}{\sqrt{p^2_r+\frac{p^2_{\theta}}{r^2}}}-K_r(r-r_c),\label{3.13}
    \\
    &&\dot{\theta}=\frac{\partial E}{\partial p_{\theta}}=\dfrac{p_{\theta}/r^2}{\sqrt{p^2_r+\frac{p^2_{\theta}}{r^2}}},\label{3.14}
    \\
    &&\dot{p_{\theta}}=-\frac{\partial E}{\partial\theta}=-K_{\theta}r_H(y-y_c).\label{3.15}
\end{eqnarray}
These are the main equations for our further numerical studies of the motion of the particle. One point must be noted here is that we've ignored the interaction between the harmonic potentials and the black hole spacetime as it is very weak compared to the other terms presented in \eqref{3.11}.

\section{Numerical findings \& analysis}\label{sec4}
In this section, we shall finally show numerically that when the system comes under the influence of the event horizon then the effect of the horizon plays a crucial role so that the radial trajectory of the particles has some instability in the context of modified gravity. Therefore, the present section will be focused on the numerical sides of our analysis i.e., the study of the Poincar$\Acute{e}$ sections of the system. 

\subsection{Analysis of the Poincar$\Acute{e}$ sections}\label{s4a}
The Poincar$\Acute{e}$ map, a profound concept within the realm of dynamical systems, is defined as the intersection of a periodic orbit residing in the expanse of state space of a continuous dynamical system, enmeshed with a lower-dimensional subspace. The essential idea is to map the higher dimensional phase-space trajectories into the lower one \cite{dalui,Dalui:2019umw}. However, the system has to satisfy some properties, namely to return to the same region in its state space from time to time. This is satisfied if the system is periodic, however, it also works with the chaotic dynamics.  

In this endeavor, for drawing the Poincar$\Acute{e}$ map takes the form of the polar plane, designated by $\theta=0$, serving as the Poincar$\Acute{e}$ section. Within this space, we plot the points on the $r-p_r$ plane when the particle gracefully intersects the Poincar$\Acute{e}$ section, guided by the constraint $p_{\theta}>0$. So, for the periodic case, the plotted points will move on a torus in the phase-space while for the chaotic scenario, some of these tori will be broken. These broken distribution of the tori will determine whether our system has become chaotic or not.

Now, we'll solve the dynamical equations of motion \big(\eqref{3.12}, \eqref{3.13}, \eqref{3.14}\big), and \eqref{3.15} for static spherically symmetric charged and charge-less neutral black hole in the context of $f(R)$ gravity by analyzing the Poincar$\Acute{e}$ sections of the dynamical systems. We've already showed analytically that the radial trajectory of the particles induces the exponential growth in the presence of horizon, which indicate a chaotic behaviour for both charged and charge-less black hole solutions in the previous section \ref{sec3}. Thus, first, we present the Poincar$\Acute{e}$ sections for the SSS charged black hole solution and then those for the SSS neutral black hole solution, in where we systematically analyse the effects of the dimensional parameter $a$ with the inclusion of horizon effect on the chaotic fluctuations.

\subsubsection{Charged black hole}\label{s4a1}
For charged SSS black hole solution \eqref{2.21}, the dynamical equations of motion \big(\eqref{3.12},
\eqref{3.13},\eqref{3.14}\big) and \eqref{3.15} are numerically solved using the fourth order Runge-Kutta method with fixed $d\tau=10^{-3}$. Here we have analyzed two types of the Poincar$\Acute{e}$ maps i.e., for different energies with fixed dimensional parameter $a$ and for fixed energy $E$ with different values of $a$. In the whole scenario, we have considered $K_r=400$, $K_{\theta}=100$, $y_c=0$, and $r_c=4.5$. The value of $p_{\theta}$ is obtained from \eqref{3.5} for a fixed value of energy under the consideration of outgoing particles only. The other variables such as $r,p_r$ and $\theta$ are initialized with the generations of the random numbers in the programming.

\begin{figure}[!ht]
	\begin{center} 
		$\begin{array}{ccc}
			\subfigure[]{\includegraphics[width=0.54\linewidth,height=0.3\linewidth]{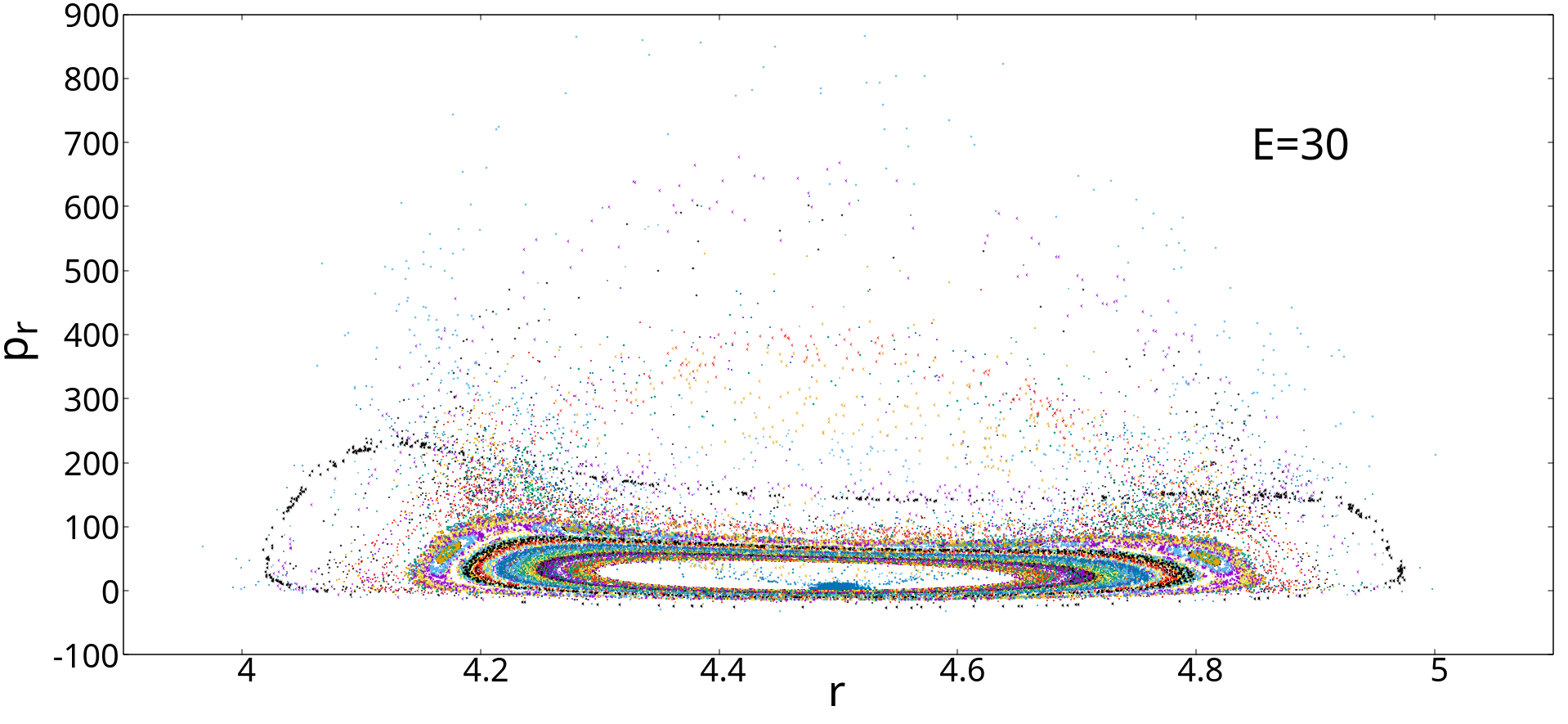}\label{3a}}
			\subfigure[]{\includegraphics[width=0.54\linewidth,height=0.3\linewidth]{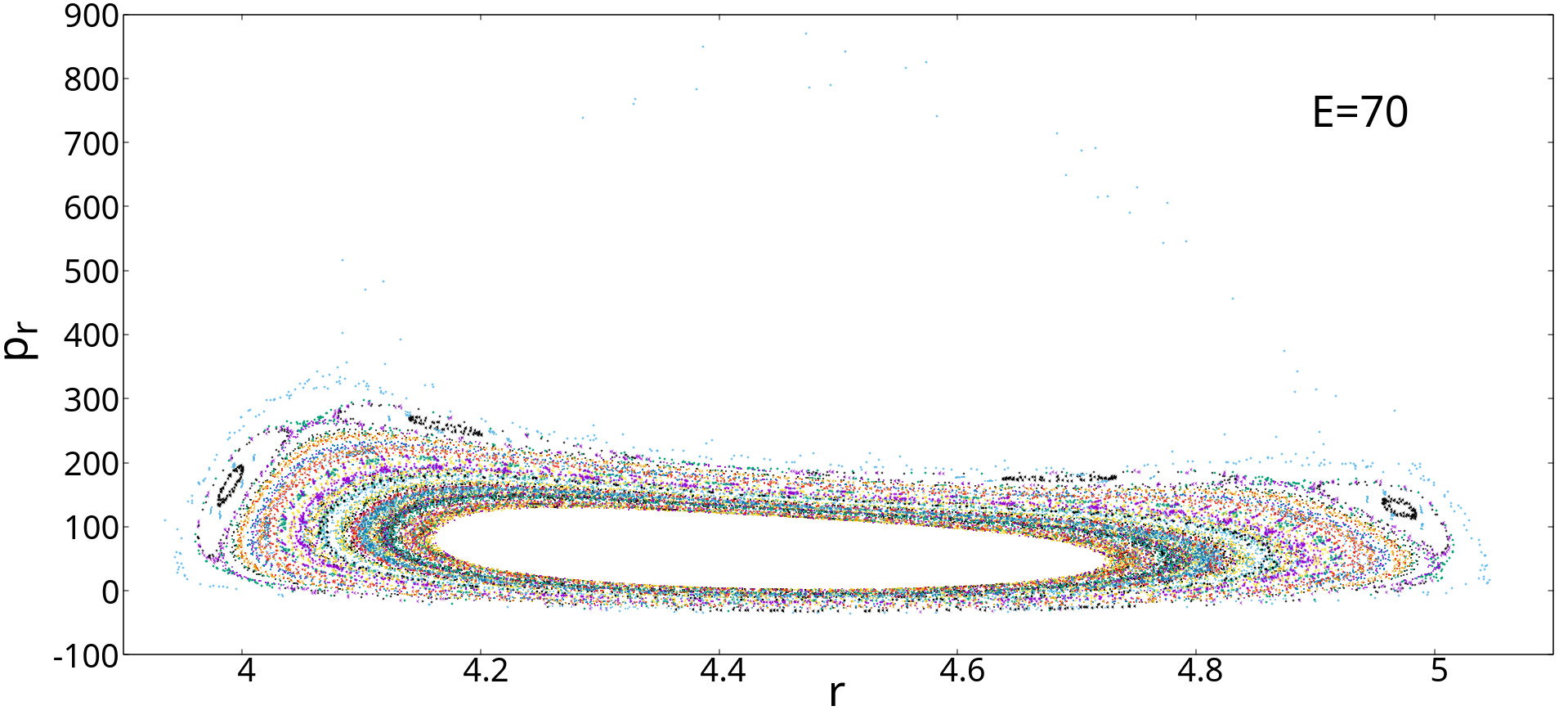}\label{3b}}\\
			\subfigure[]{\includegraphics[width=0.54\linewidth,height=0.3\linewidth]{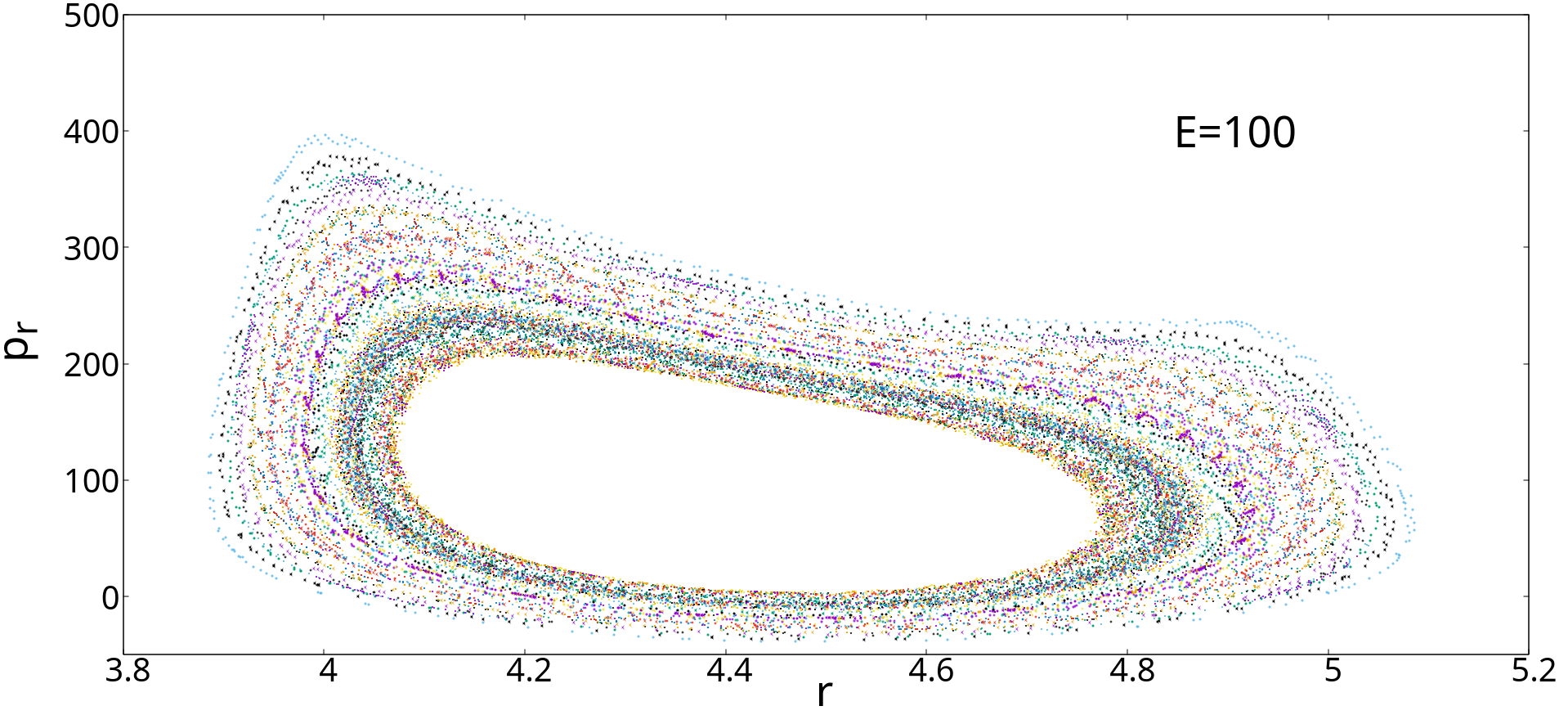}\label{3c}}
			\subfigure[]{\includegraphics[width=0.54\linewidth,height=0.3\linewidth]{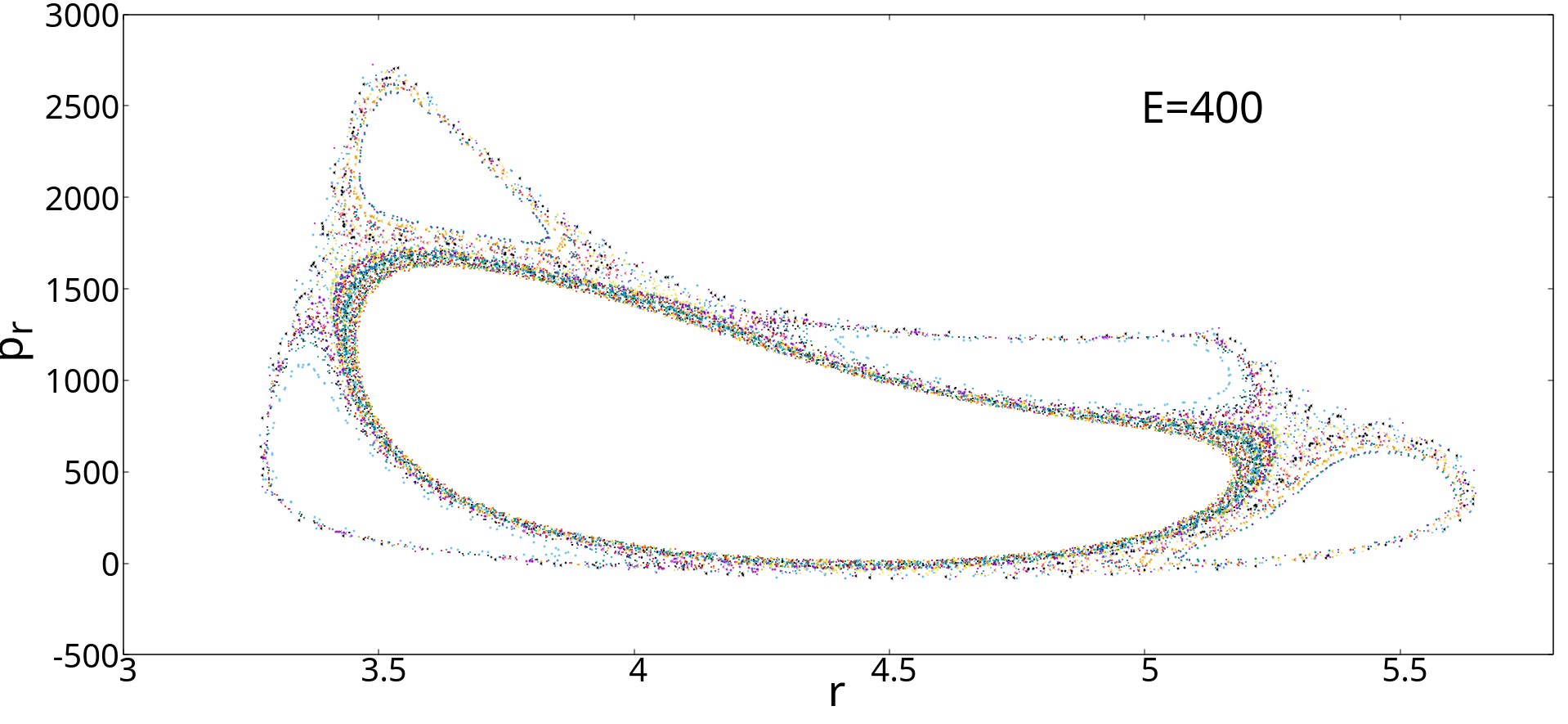}\label{3d}}\\
			\subfigure[]{\includegraphics[width=0.54\linewidth,height=0.3\linewidth]{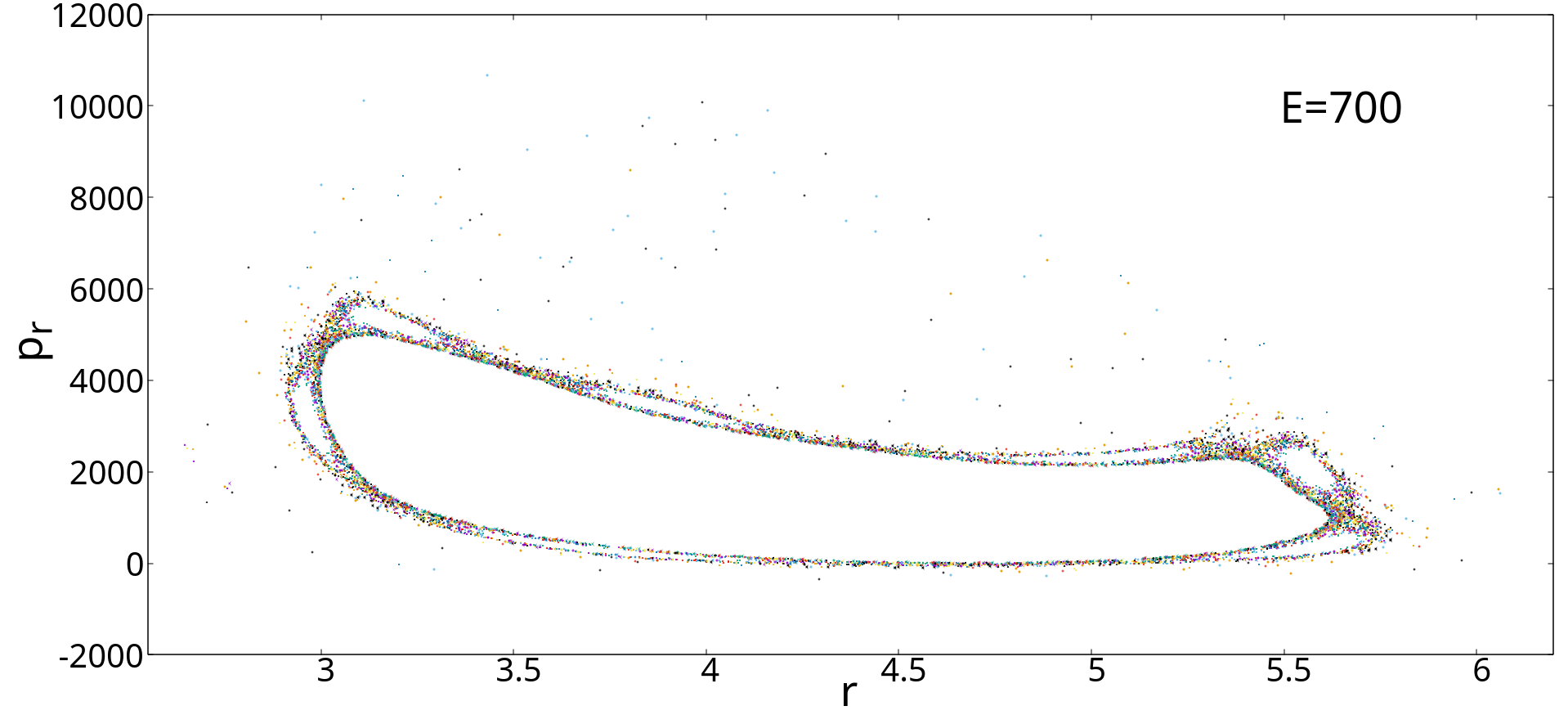}\label{3e}}
			\subfigure[]{\includegraphics[width=0.54\linewidth,height=0.3\linewidth]{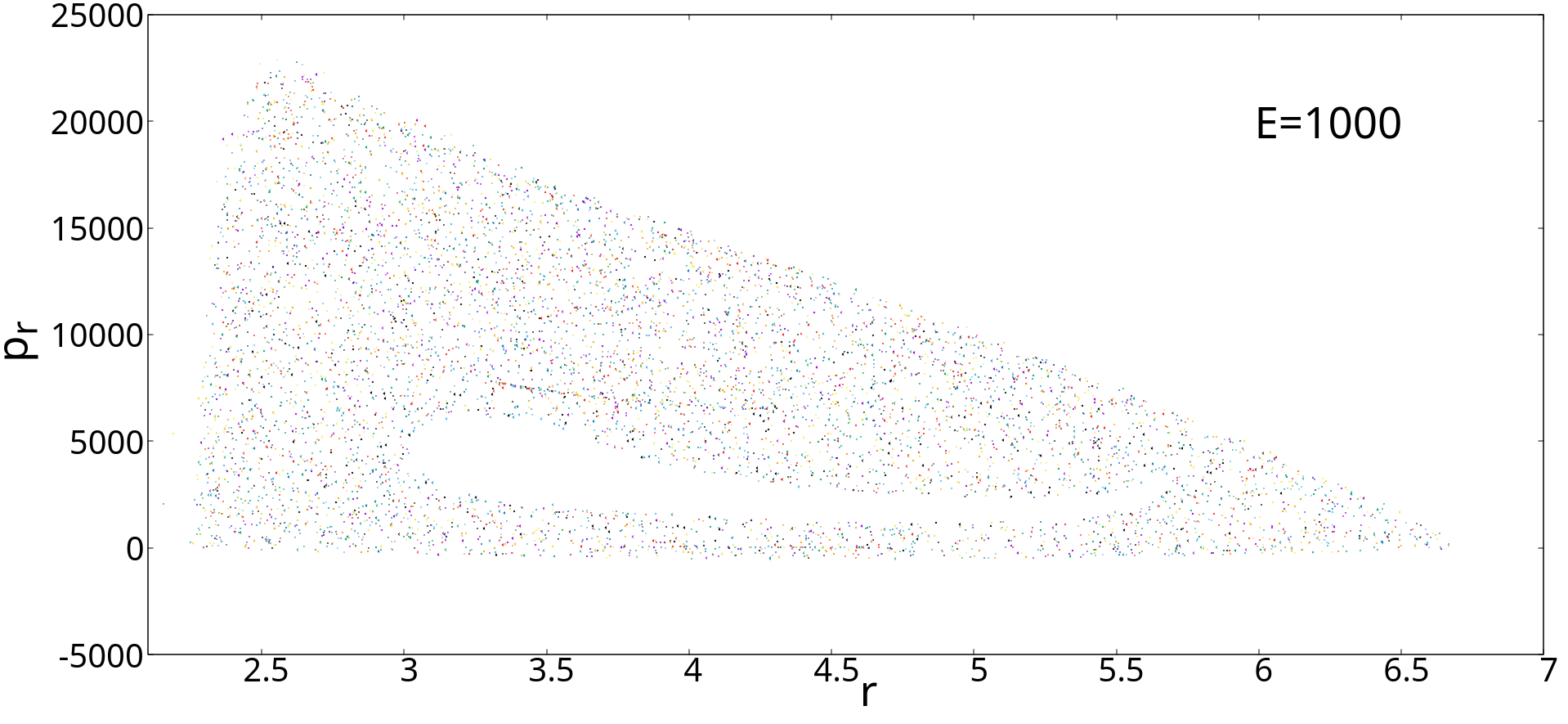}\label{3f}}
		\end{array}
		$\end{center}
	\caption{The Poincar$\Acute{e}$ sections in the $(r-p_r)$ plane with $p_{\theta}>0$ and $\theta=0$ with different energies with fixed dimensional parameter $a=0.166$ for the SSS charged black hole.}
	\label{f3}
\end{figure}

As we have already discussed about the upper bound of the value of dimensional parameter $a$ as $0.166$ in the subsection \ref{s2b}, here we have considered different energies such as $E=30,70,100,400,700$ and $E=1000$ with a fixed value of $a$ parameter as $a=0.166$. In Fig.(\ref{f3}), we show the Poincar$\Acute{e}$ sections of the radial outgoing particles trajectory which is projected over the $r-p_r$ phase plane for the charged SSS black hole with these different energies, constrained by $\theta=0$ and $p_{\theta}>0$. Now it is clearly visible from Fig(\ref{3a}) that for the low energy $E=30$, the Poincar$\Acute{e}$ sections exhibits a regular tori so that the corresponding orbit is mainly confined near the center of the harmonic potential, which is taken as $r_c=4.5$. Similar kind of tori is also observed from Figs.(\ref{3b},\ref{3c}) for $E=70$ and $100$, respectively. These particular tori is know as Kolmogorov-Arnold-Moser (KAM) tori \cite{KAM}. But due to the conservation of the Hamiltonian of the system, as the total energy of the system is increased, the momentum of the system will also be increased and as a consequence the trajectory of the particle system approaches near the black hole event horizon $r_{+}=2.135$, which is fixed due to fixed dimensional parameter $a$. As a result for $E=400$ and $E=700$, the Poincar$\Acute{e}$ section shows that KAM tori starts getting distorted and appeared to be pinched, as shown in the Figs.(\ref{3d},\ref{3e}), respectively. Finally from Fig.(\ref{3f}), it is easily observed that further increasing in energy as $E=1000$ makes the complete breaking of the regular tori and as a results the points are distributed in the phase plane. From the complete visualization of the Fig(\ref{f3}), it is noteworthy to mention that with increasing the value of energy makes the center of the corresponding orbits shifted towards the event horizon besides breaking the regular tori. Therefore in this analysis, we have observed that the radial coordinate $r$ becomes less than the event horizon radius $r_{+}$ during time evolution makes numerical instabilities in our programming calculations for increasing the energy more than above $E=1000$. so that, we have some upper bound of the energy due to the effect of the horizon. This particular features of the Poincar$\Acute{e}$ section supports the chaotic nature of the particle trajectory near the horizon. Here different colors in the figures indicate the trajectory of the particles are solved for the different initial conditions.


\begin{figure}[H]
	\begin{center} 
		$\begin{array}{ccc}
			\subfigure[]{\includegraphics[width=0.54\linewidth,height=0.3\linewidth]{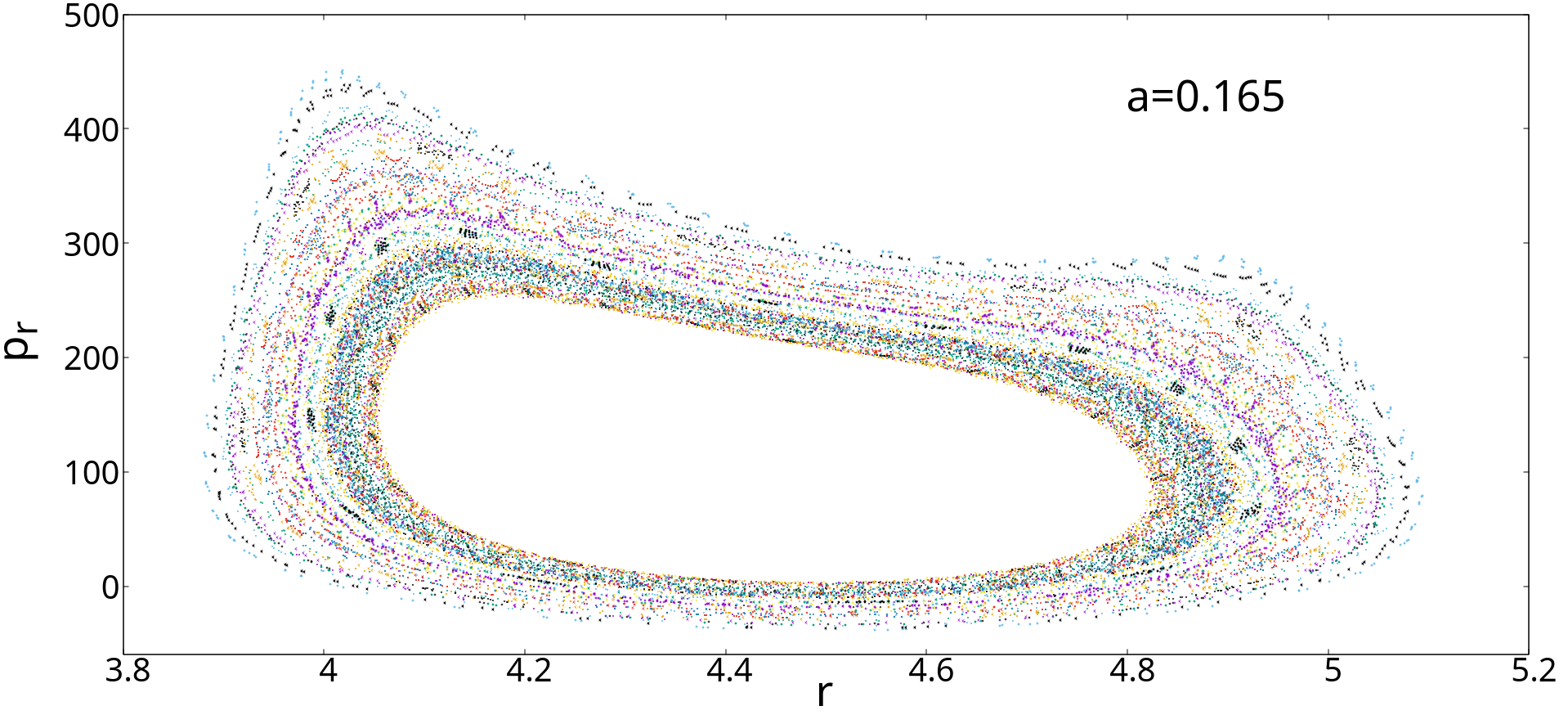}\label{4a}}
			\subfigure[]{\includegraphics[width=0.54\linewidth,height=0.3\linewidth]{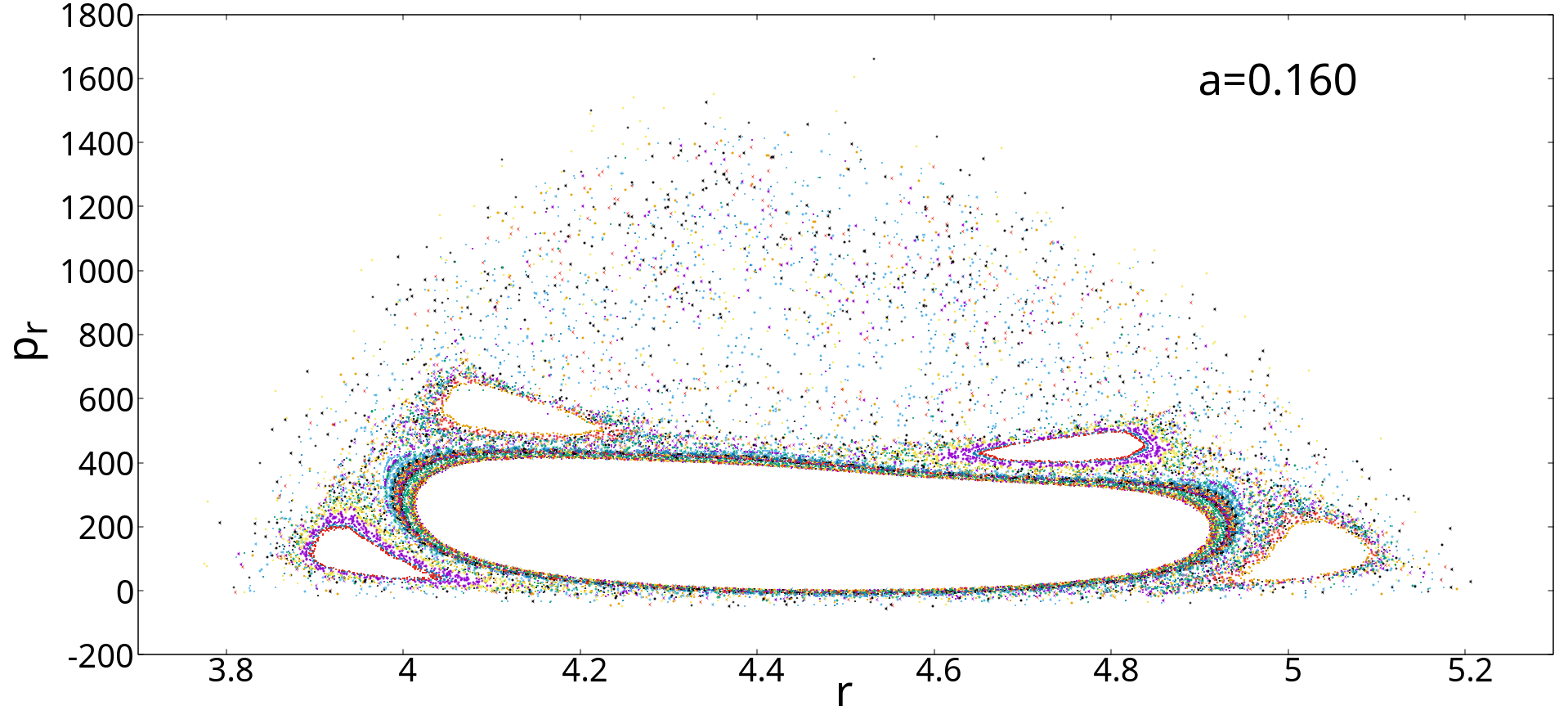}\label{4b}}\\
			\subfigure[]{\includegraphics[width=0.54\linewidth,height=0.3\linewidth]{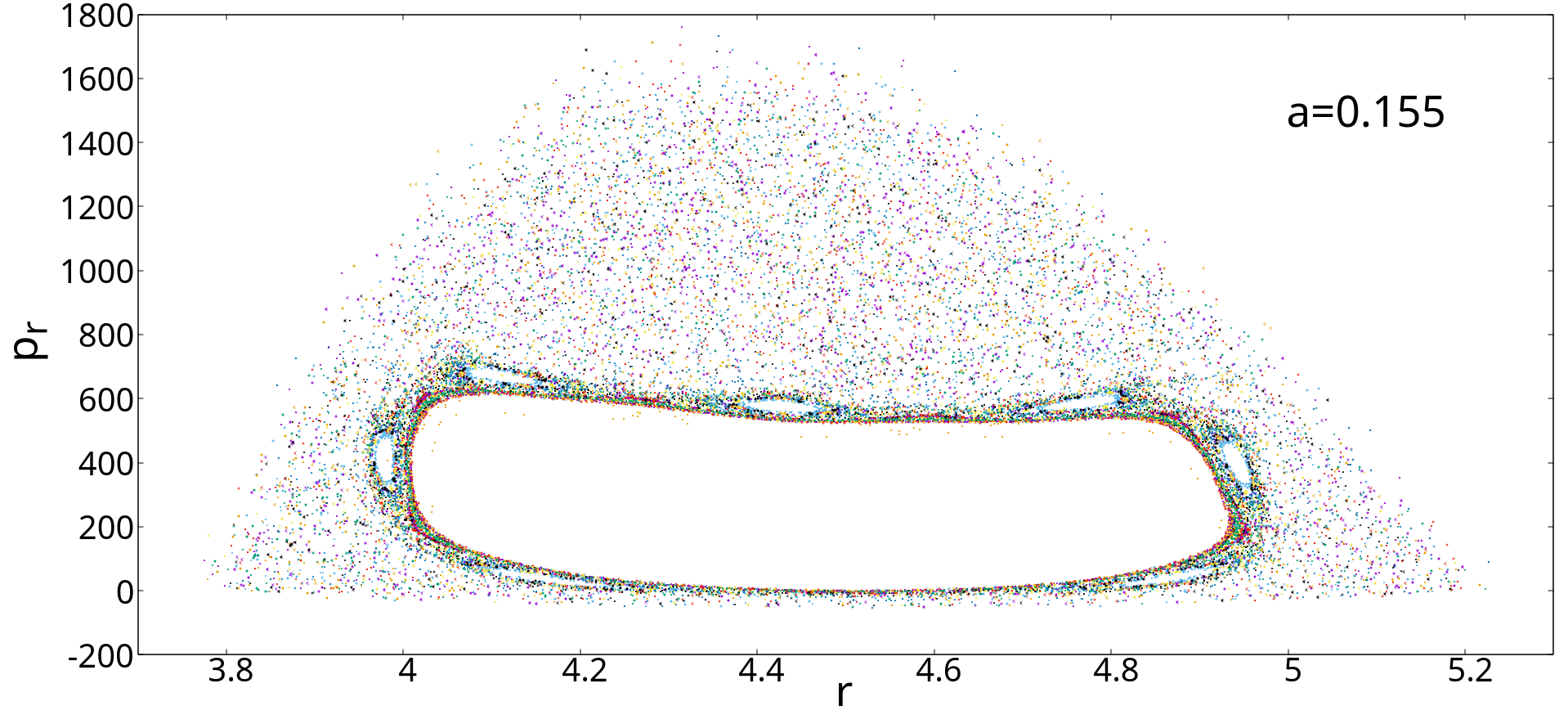}\label{4c}}
			\subfigure[]{\includegraphics[width=0.54\linewidth,height=0.3\linewidth]{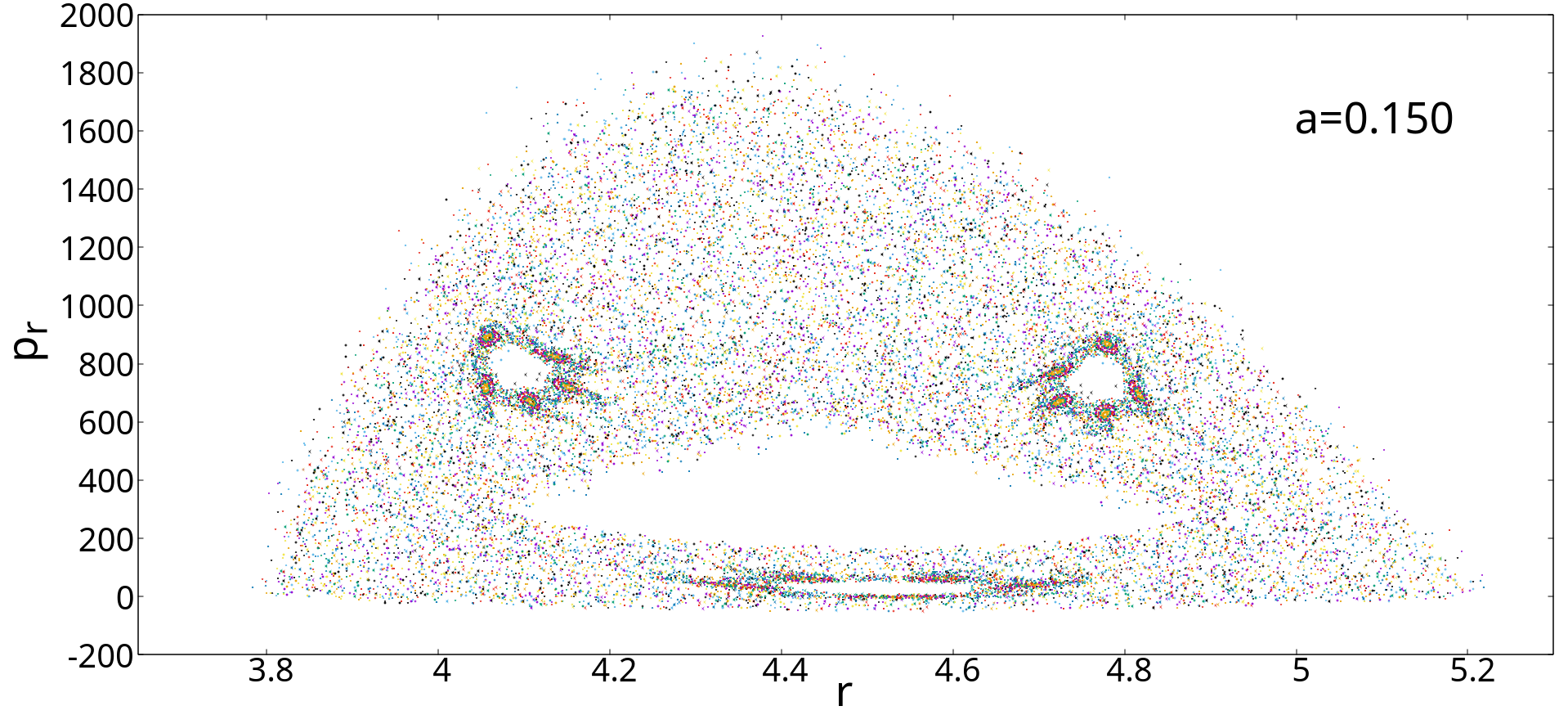}\label{4d}}\\
			\subfigure[]{\includegraphics[width=0.54\linewidth,height=0.3\linewidth]{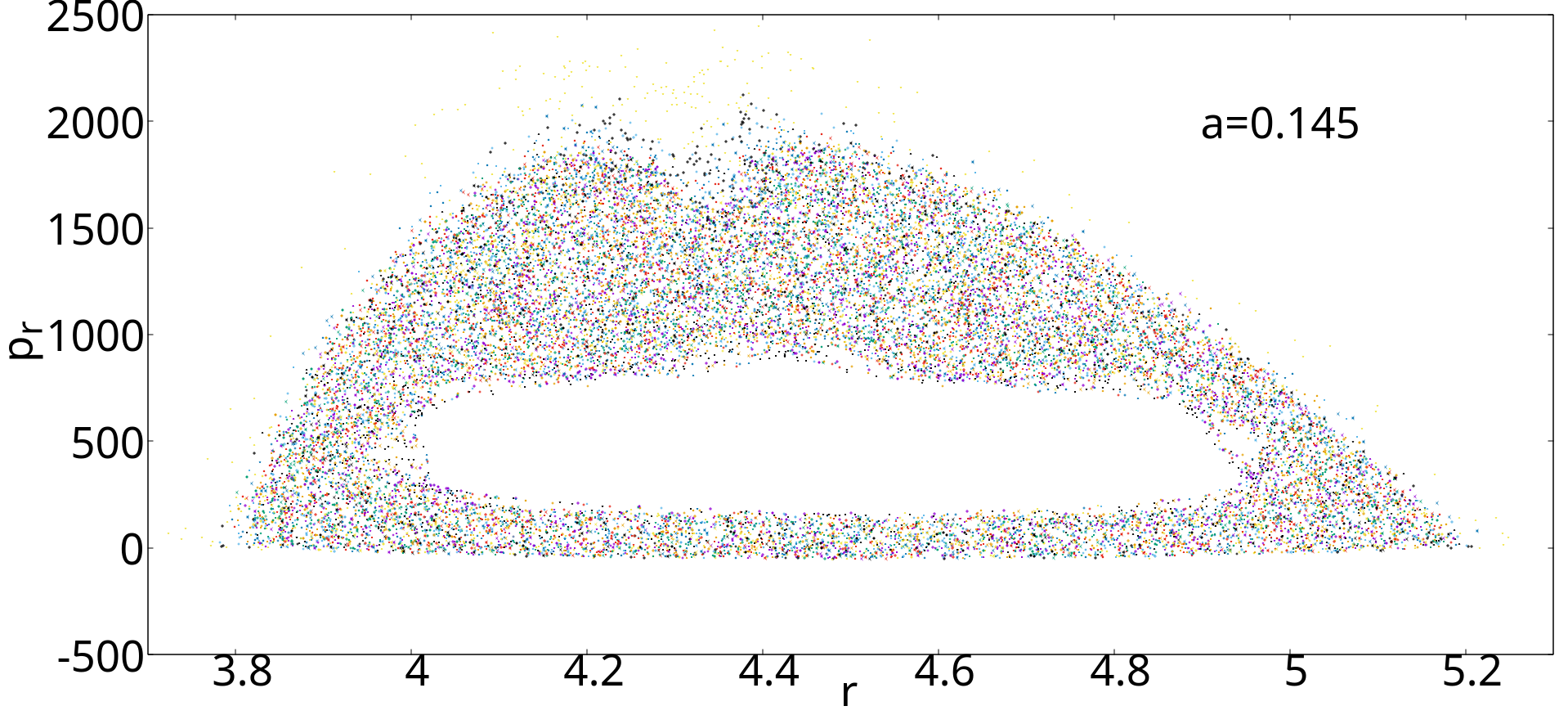}\label{4e}}
			\subfigure[]{\includegraphics[width=0.54\linewidth,height=0.3\linewidth]{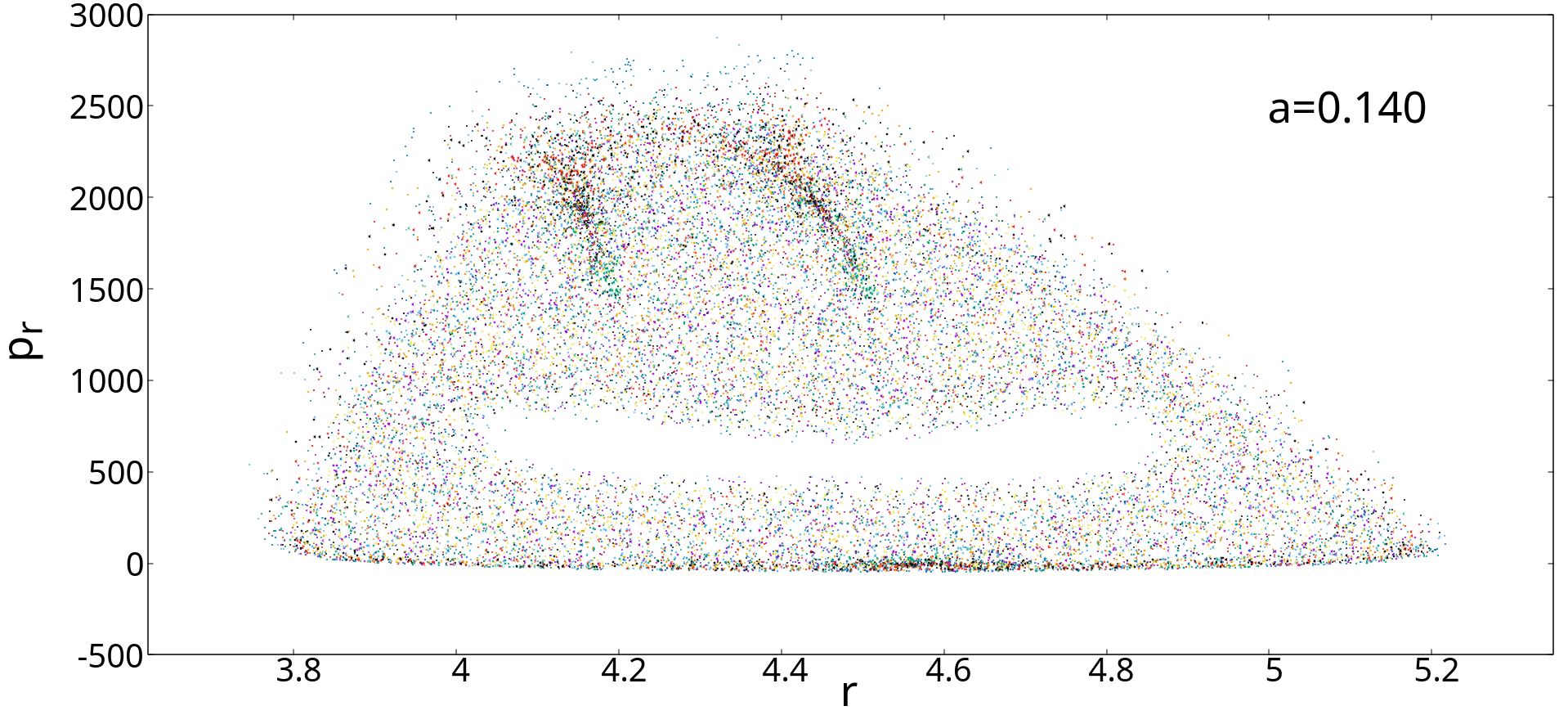}\label{4f}}\\
			\subfigure[]{\includegraphics[width=0.54\linewidth,height=0.3\linewidth]{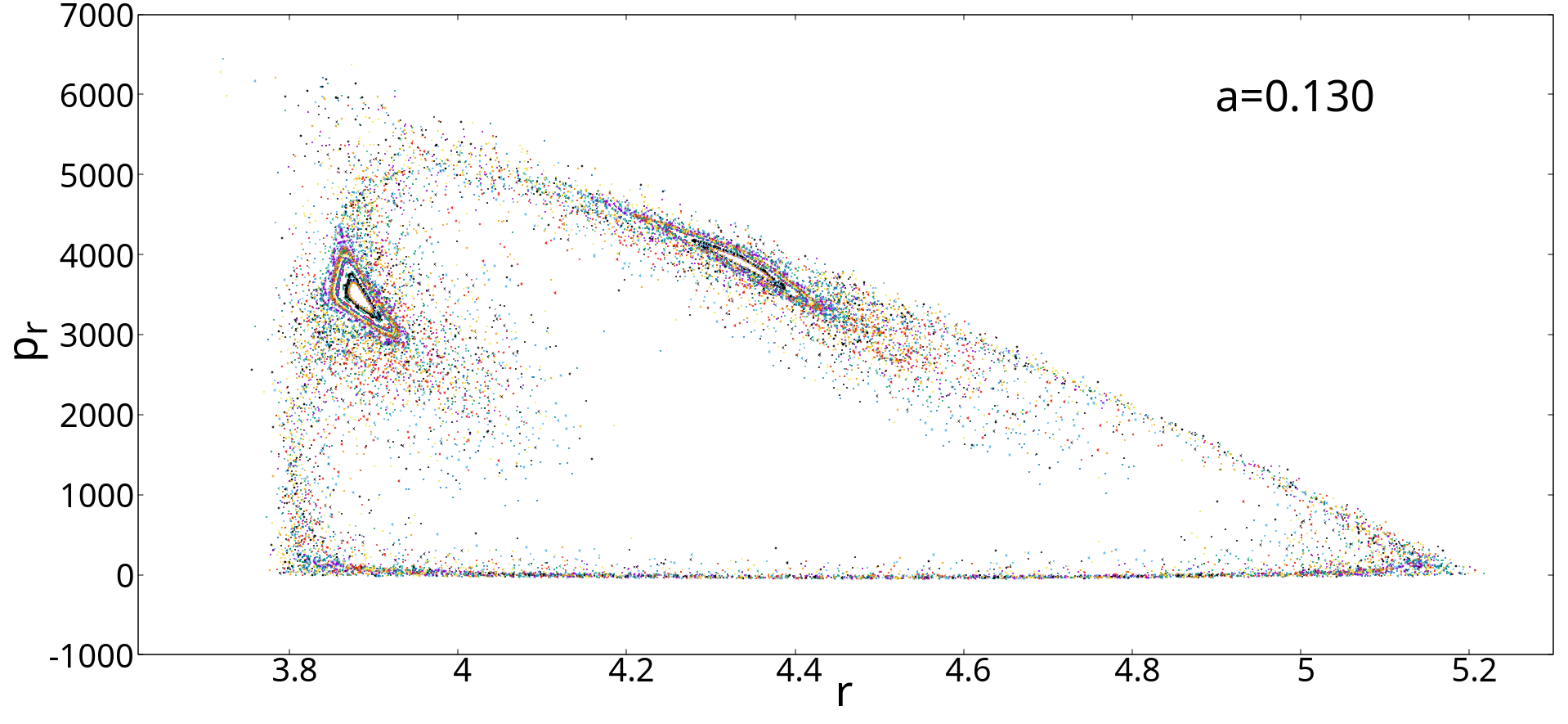}\label{4g}}
		\end{array}
		$\end{center}
	\caption{The Poincar$\Acute{e}$ sections in the $(r-p_r)$ plane with $p_{\theta}>0$ and $\theta=0$ with different dimensional parameter $a$ for fixed energy $E=100$ for the SSS charged black hole.}
	\label{f4}
\end{figure}

In the next part, we have analyzed the nature of the radial outgoing particle trajectory for a fixed value of energy $E=100$ by changing the dimensional parameter $a$. Therefore in Fig.(\ref{f4}), we plot the different Poincar$\Acute{e}$ sections for the different dimensional parameter $a=0.165,0.160,0.155,0.150,0.145,0.140,0.130$ with a fixed energy in the $r-p_r$ phase plane. For $a=0.165$ with $E=100$, it makes a perfect KAM tori \big(see Fig.(\ref{4a}\big). Then with decreasing $a$, the regular tori started breaking and for $a=0.145$ the tori completely breaks down. Therefore, in this case with the variation of the parameter $a$ the radius of the horizon starts varying and with the decreasing value of $a$ the horizon radius increases (from Fig. (\ref{f1})) and it approaches near to the particle with this variation. As a result, the particle system starts interacting with the horizon and the total dynamical system becomes chaotic. 
It is also clear from Fig.(\ref{f4}) that initially for $a=0.165$, the maximum value of the radial component of the linear momentum is 500 but when the parameter $a$ is decreased, it becomes almost $3000$ for $a=0.140$, which means that the system represents a chaotic fluctuations when it interacts with the horizon. Finally it is observed from Fig.(\ref{4g}) that, for $a=0.130$, points are distributed in the phase plane means most of the particles are already disappeared by the effect of the horizon in the context of modified gravity. Hence, we clearly find that the decrease in parameter $a$ introduces chaotic fluctuations in the trajectories, and at very low value of $a$ (likely for $a=0.140,0.130$) the particle trajectory becomes fully chaotic in nature. It should be mentioned that the introduction of the dimensional parameter $a$, signifies the effects of modified gravity i.e. $f(R)$ gravity.

\subsubsection{Neutral black hole}\label{s4a2}
\begin{figure}[H]
	\begin{center} 
		$\begin{array}{ccc}
			\subfigure[]{\includegraphics[width=0.54\linewidth,height=0.3\linewidth]{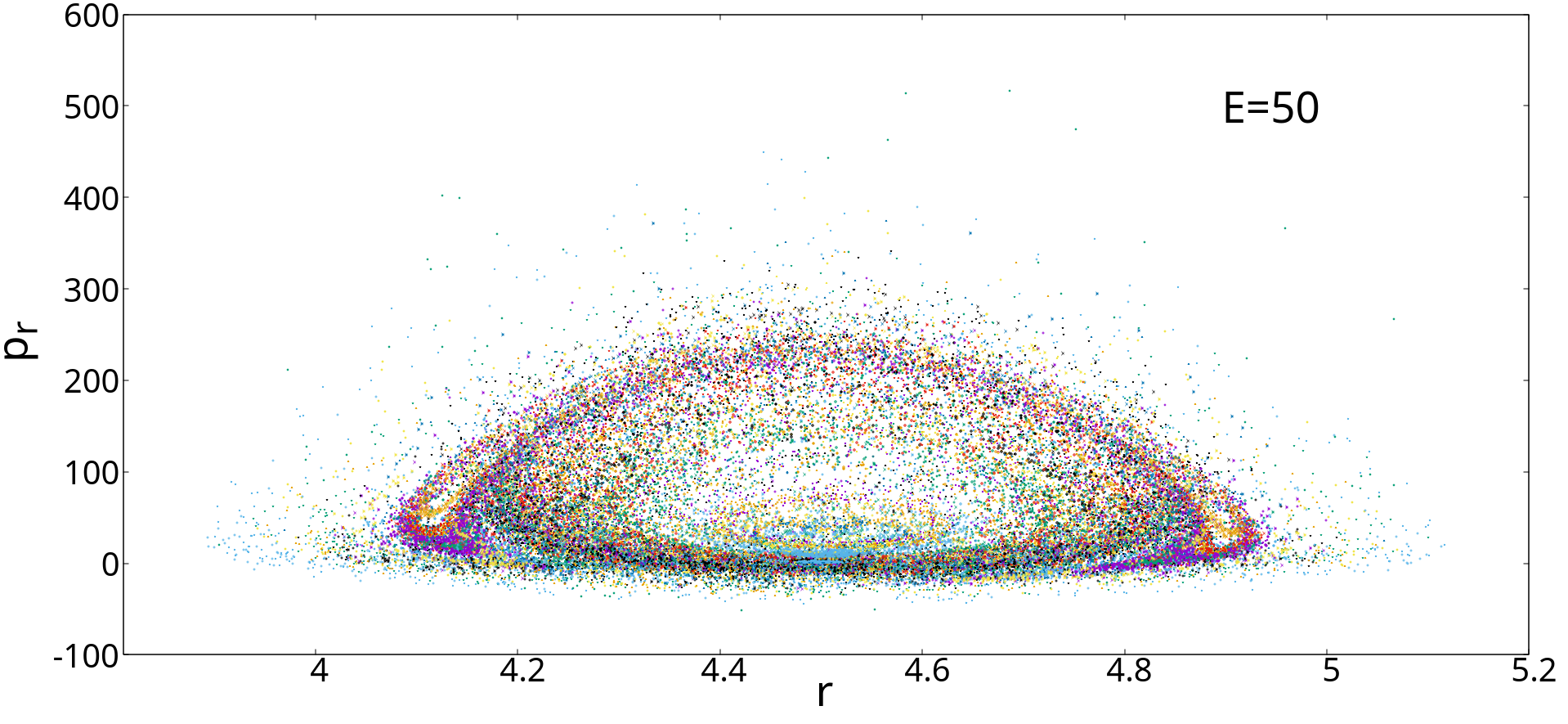}\label{5a}}
			\subfigure[]{\includegraphics[width=0.54\linewidth,height=0.3\linewidth]{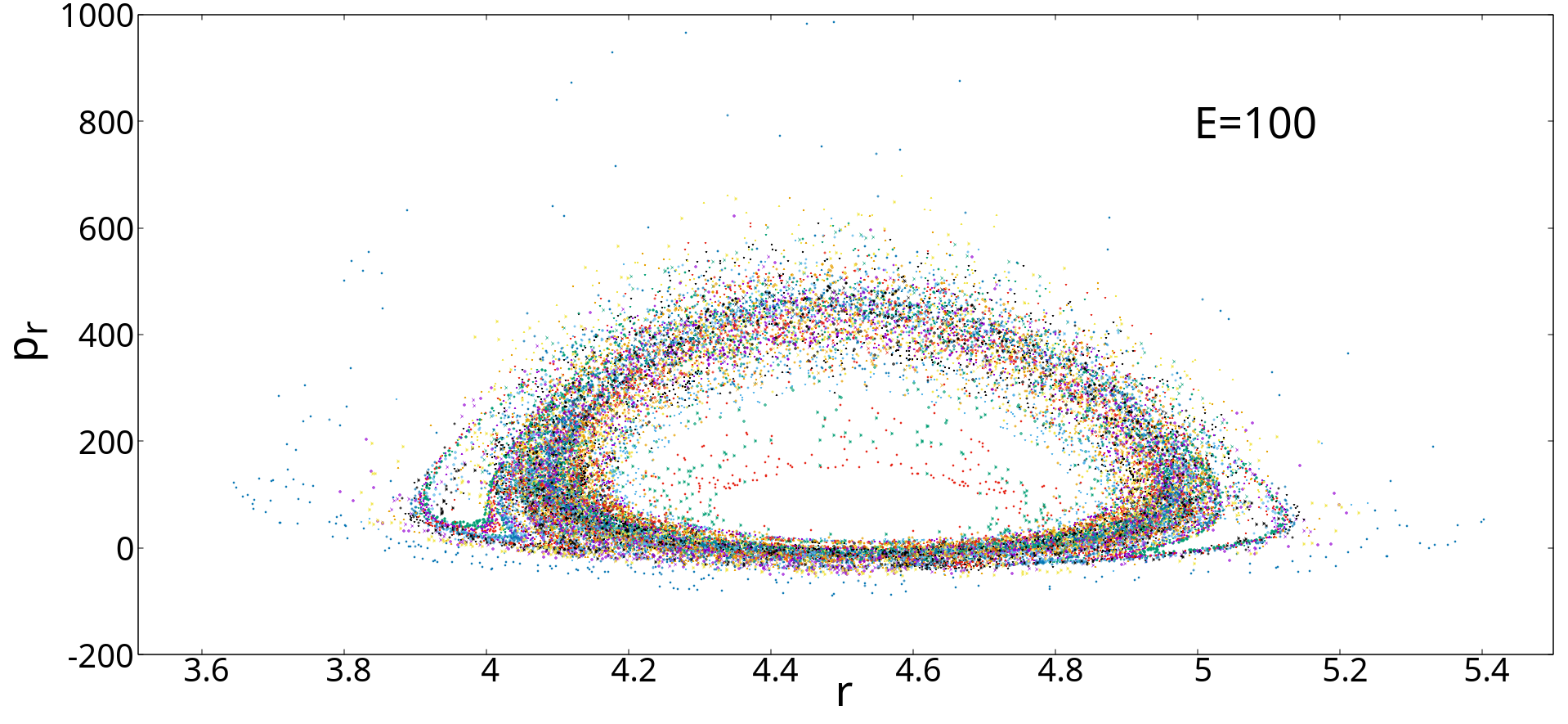}\label{5b}}\\
			\subfigure[]{\includegraphics[width=0.54\linewidth,height=0.3\linewidth]{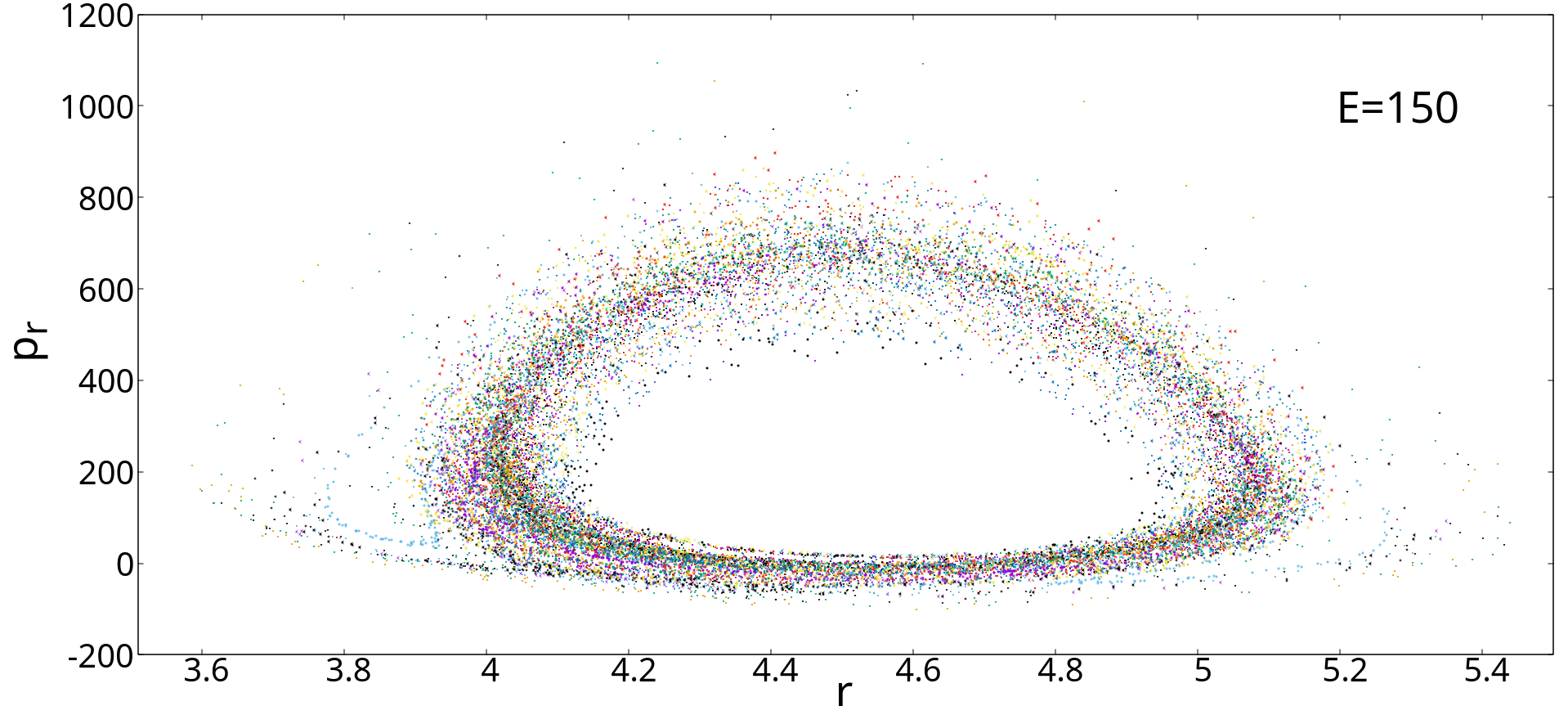}\label{5c}}
			\subfigure[]{\includegraphics[width=0.54\linewidth,height=0.3\linewidth]{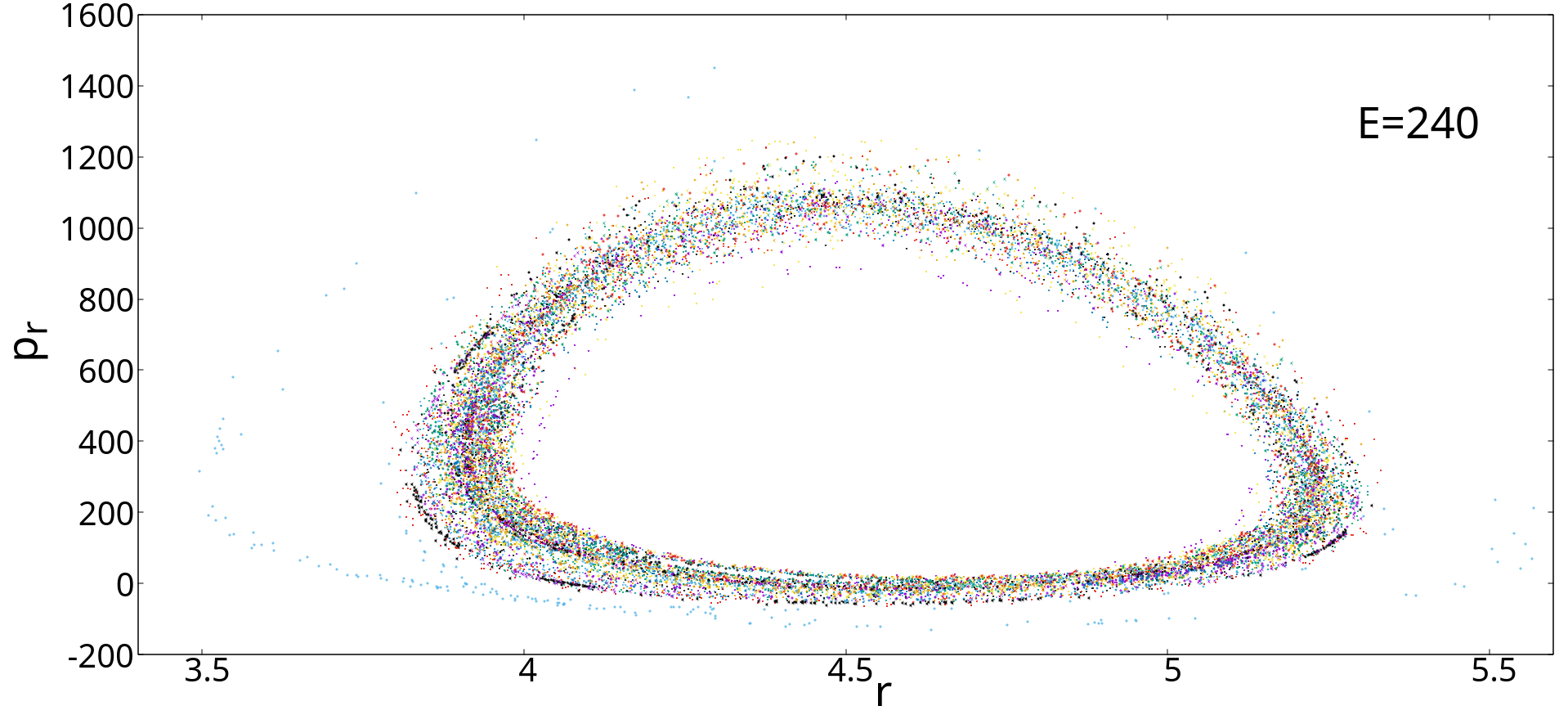}\label{5d}}\\
			\subfigure[]{\includegraphics[width=0.54\linewidth,height=0.3\linewidth]{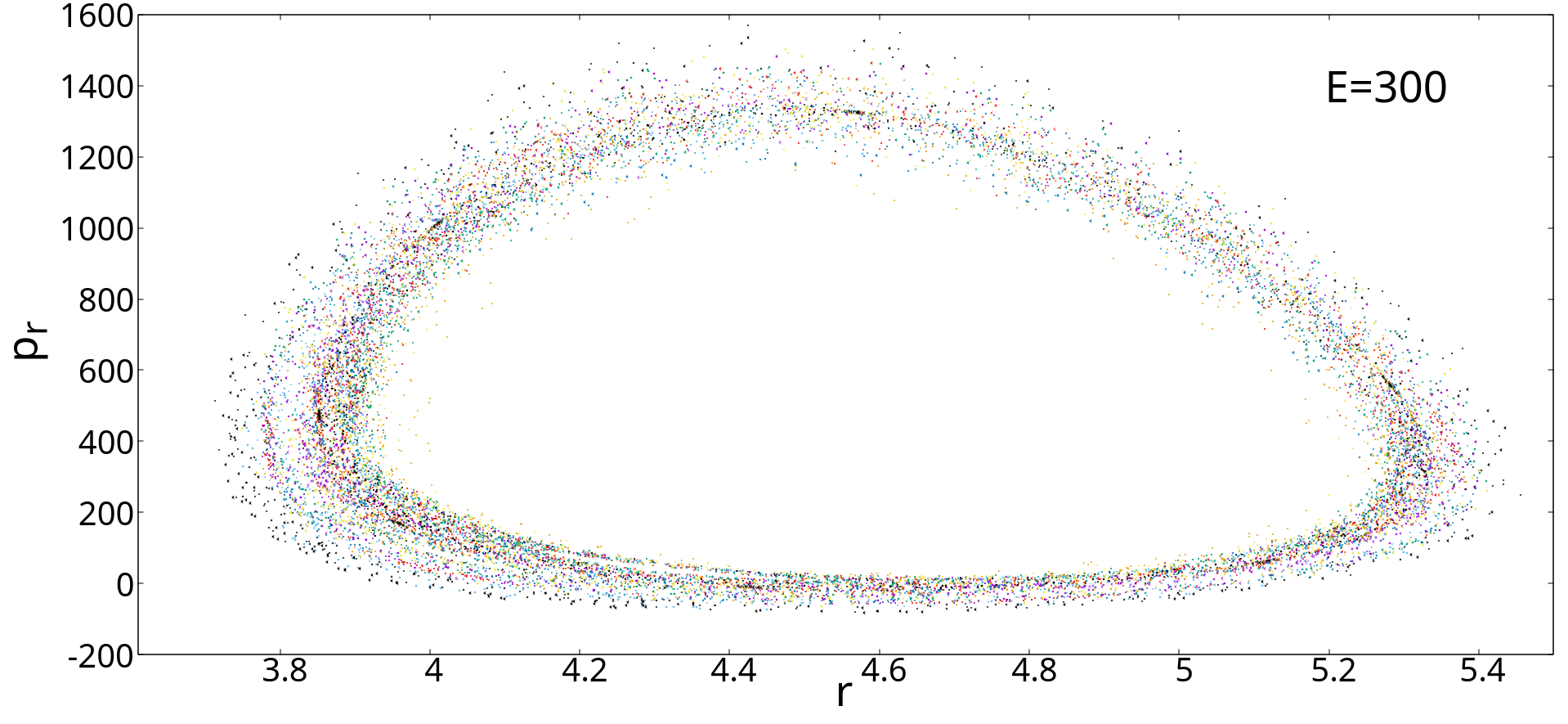}\label{5e}}
			\subfigure[]{\includegraphics[width=0.54\linewidth,height=0.3\linewidth]{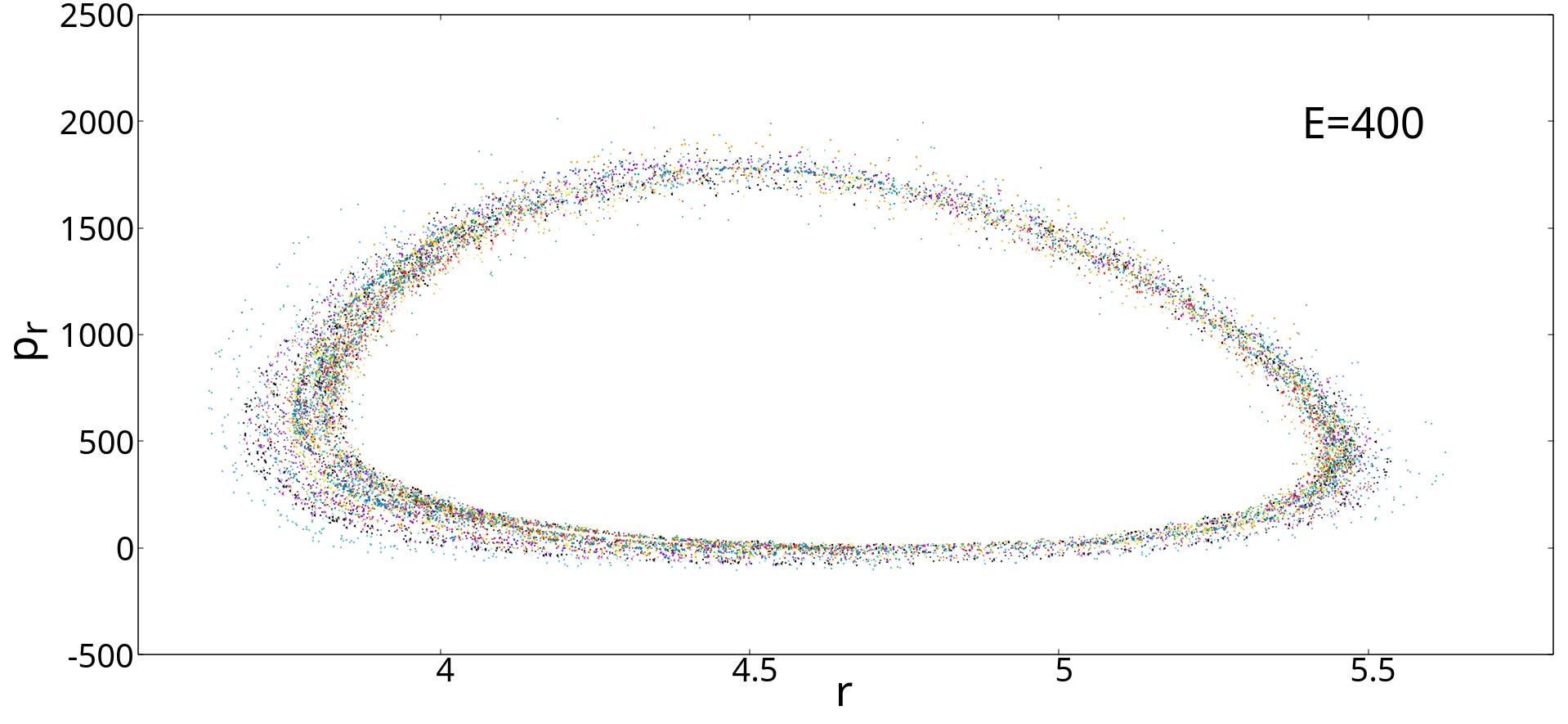}\label{5f}}
		\end{array}
		$\end{center}
	\caption{The Poincar$\Acute{e}$ sections in the $(r-p_r)$ plane with $p_{\theta}>0$ and $\theta=0$ with different energies with fixed dimensional parameter $a=0.5$ for the SSS neutral black hole.}
	\label{f5}
\end{figure}

Now, we will discuss about the horizon effect and chaotic representations for the SSS charge-less or neutral black hole solution. For neutral black hole solution \eqref{2.23}, there is no restriction for the upper bound of the dimensional parameter $a$ as the event horizon radius is given by $r_H=\frac{2}{3a},~a>0$. For the higher values of $a$, the size of the black hole will be small. All the corresponding dynamical equations of motion \big(see \eqref{3.12}, \eqref{3.13}, \eqref{3.14} and \eqref{3.15}\big) are solved using the Runge-Kutta fourth-order scheme. The initial conditions are chosen in a similar way as discussed for the SSS charged black hole model i.e., $K_r=400$, $K_{\theta}=100$, $y_c=0$, and $r_c=4.5$. All of the other parameters are considered the same as in the charged case. First in Fig.(\ref{f5}), we show all of the Poincar$\Acute{e}$ sections for different energies with fixed dimensional parameter $a$, which is chosen as $a=0.5$. In Fig.(\ref{5a}), we observe that a regular closed curves appeared at low energy $E=50$, which gets squeezed along with the appearance of region filled with the scattered points but as the energy is increased to $E=100$, some region vanished with the scattered points. As a result, upon increasing energies, the regular orbits getting started to break. Therefore at very high energy $E=400$ (Fig.(\ref{5f})), the surface of the radial outgoing trajectory approaches near the event horizon. Hence, there is a complete breaking of these closed curves which is quite evident with the filling of the region with the scattered points in the phase plane. Due to the appearances of the chaotic nature in the dynamic motion of this system, it is clearly observed that the radial components of the momentum of particles increases along with increasing energy which is similar, as discussed in the above case. It should also be mentioned here that, in the case of charged SSS black hole, the chaotic nature appears at relatively higher energy ($E=1000$), but in this charge-less case it appears at relatively lower energy ($E=400$). This particular peculiar feature confirms that the term $\frac{1}{3ar^2}$, which is basically responsible for the charge plays an important role for the chrged black hole solution. On the other hand, the dimensional parameter $a$ has a very interesting role for playing with the introduction as well as the reconstruction for modified gravity as $f(R)$ gravity. The parameter $a$ of the black hole introduces more chaotic fluctuations in the radial outgoing trajectories of the particles when they are interacting with the horizon.

\begin{figure}[ht!]
	\begin{center} 
		$\begin{array}{ccc}
			\subfigure[]{\includegraphics[width=0.54\linewidth,height=0.3\linewidth]{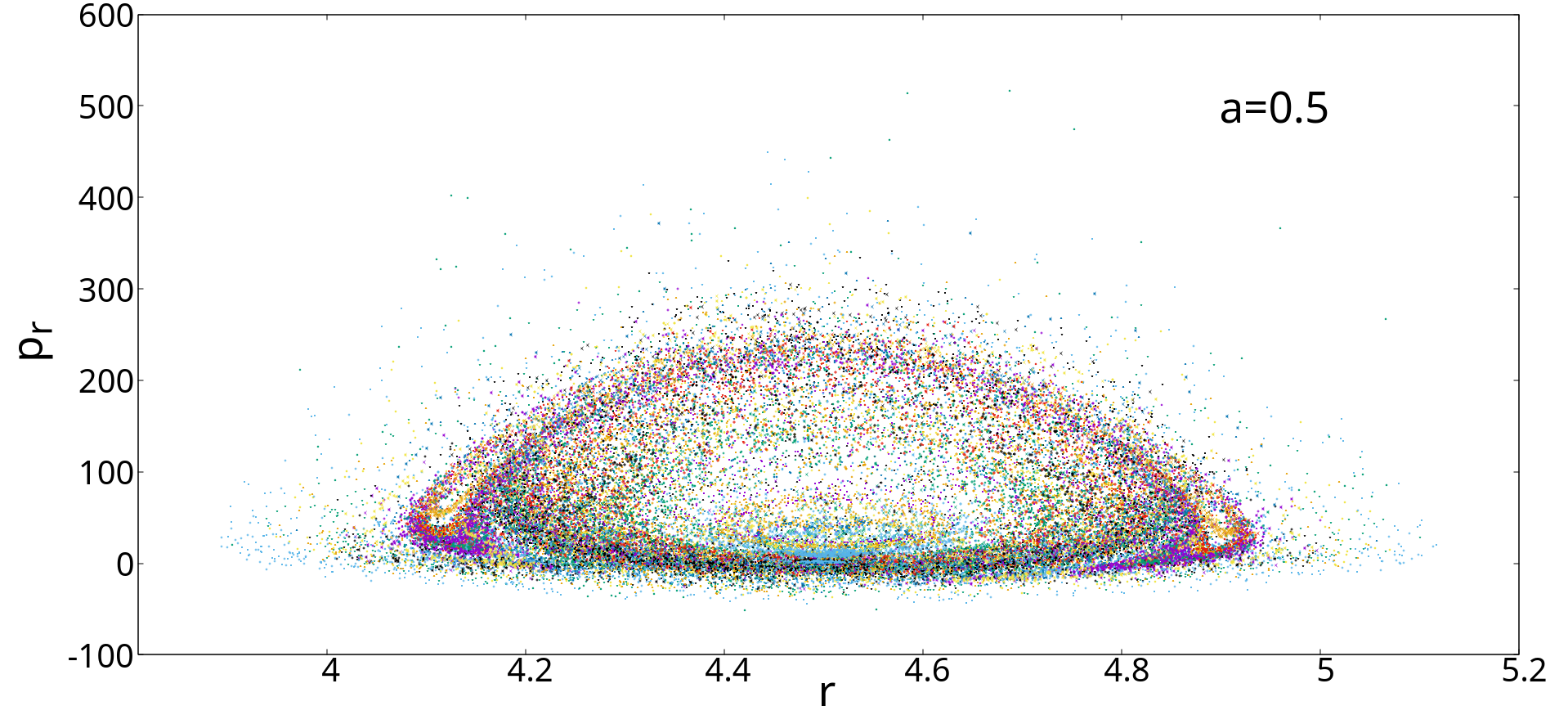}\label{6a}}
			\subfigure[]{\includegraphics[width=0.54\linewidth,height=0.3\linewidth]{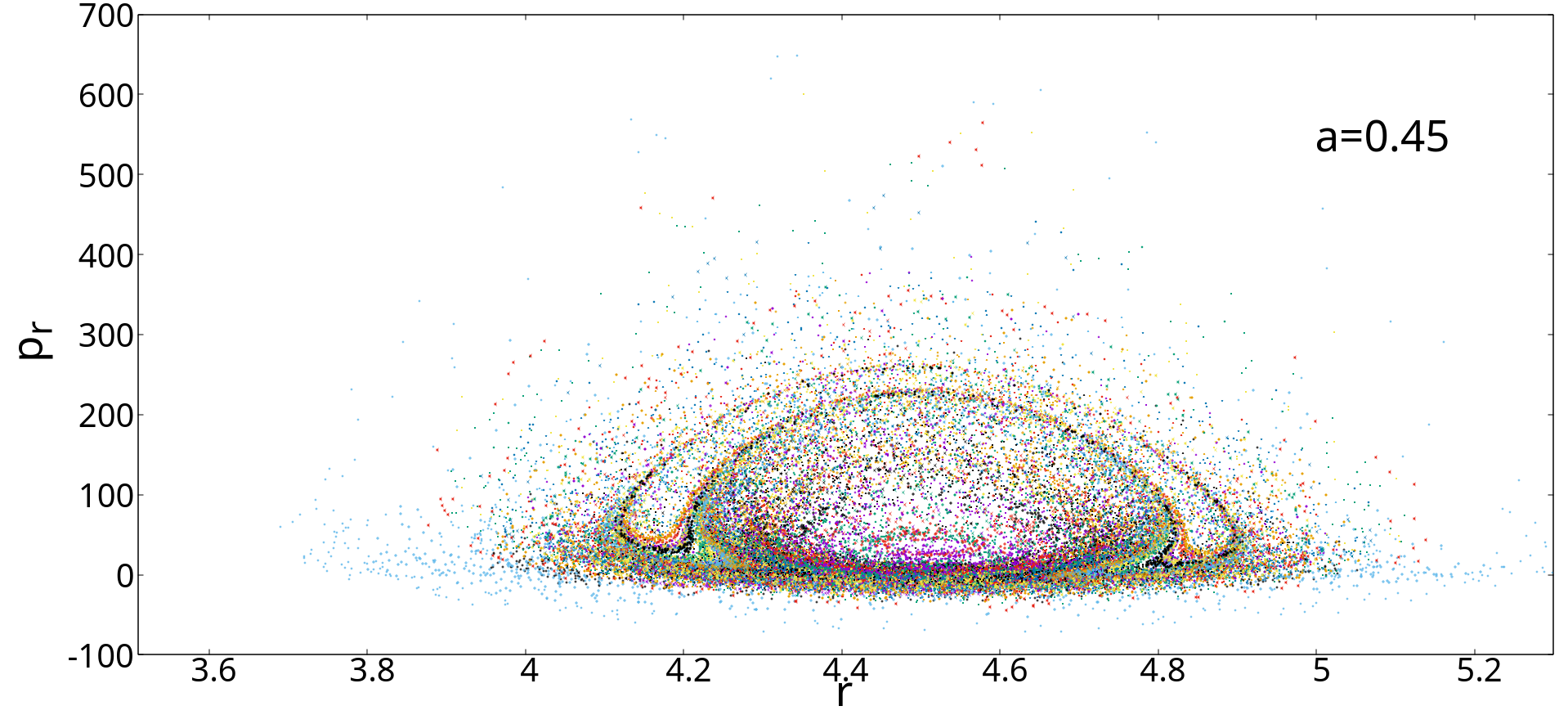}\label{6b}}\\
			\subfigure[]{\includegraphics[width=0.54\linewidth,height=0.3\linewidth]{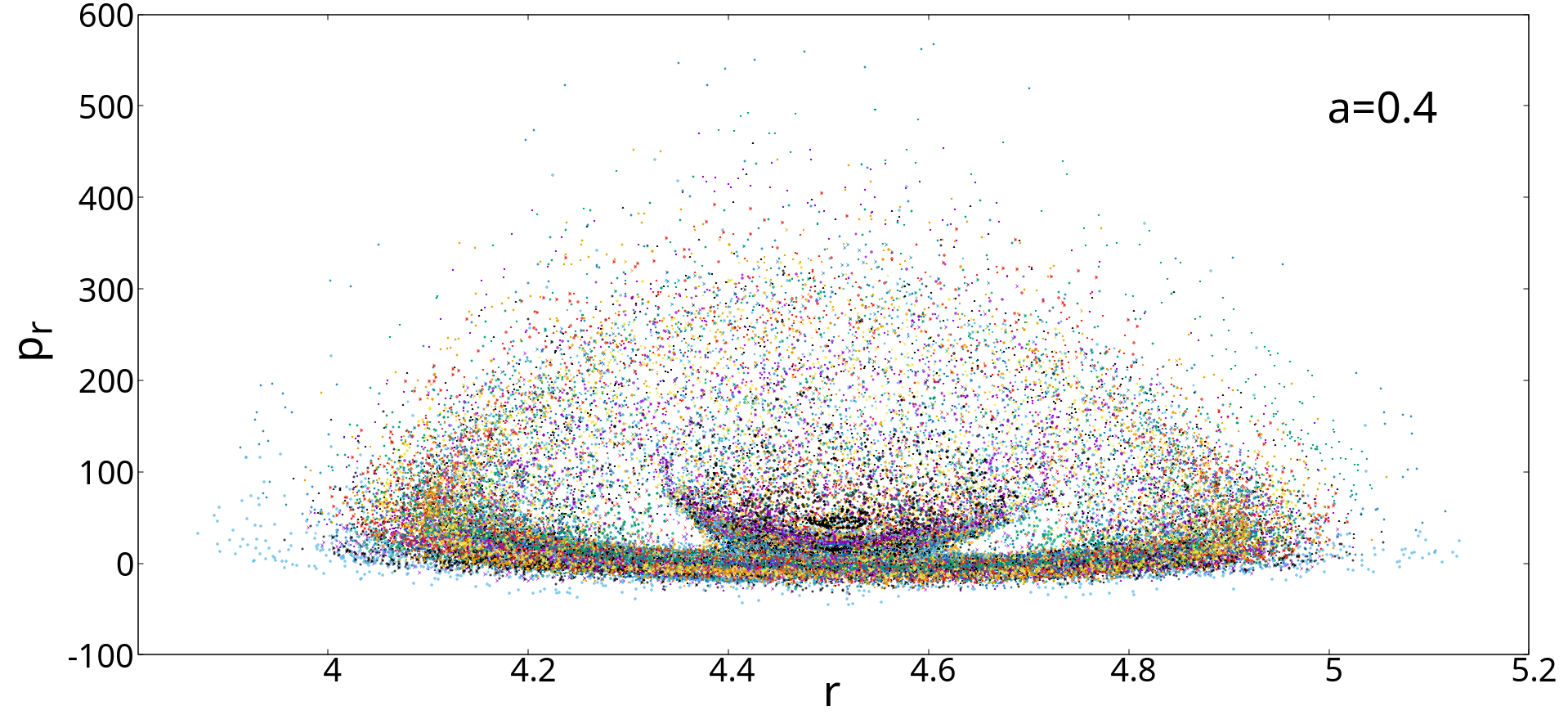}\label{6c}}
			\subfigure[]{\includegraphics[width=0.54\linewidth,height=0.3\linewidth]{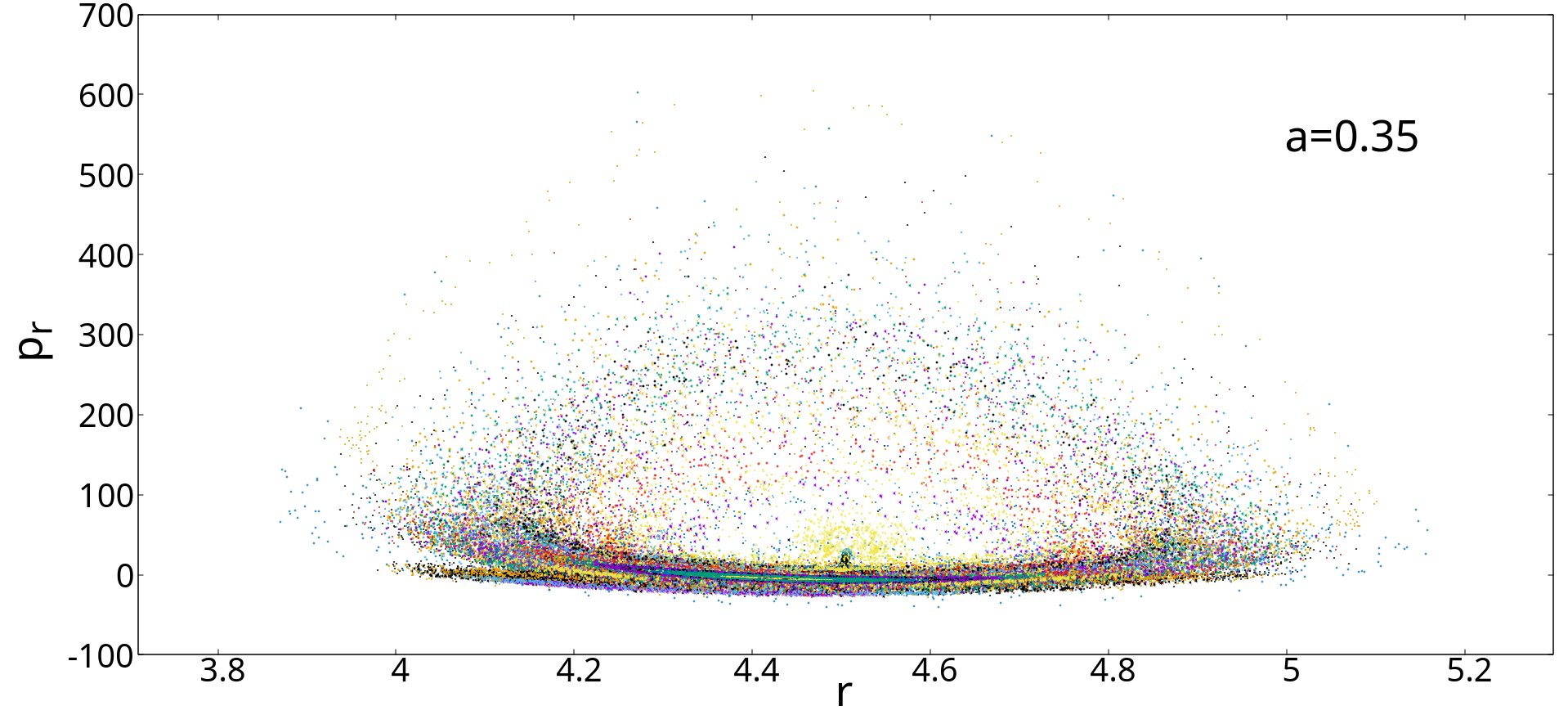}\label{6d}}\\
			\subfigure[]{\includegraphics[width=0.54\linewidth,height=0.3\linewidth]{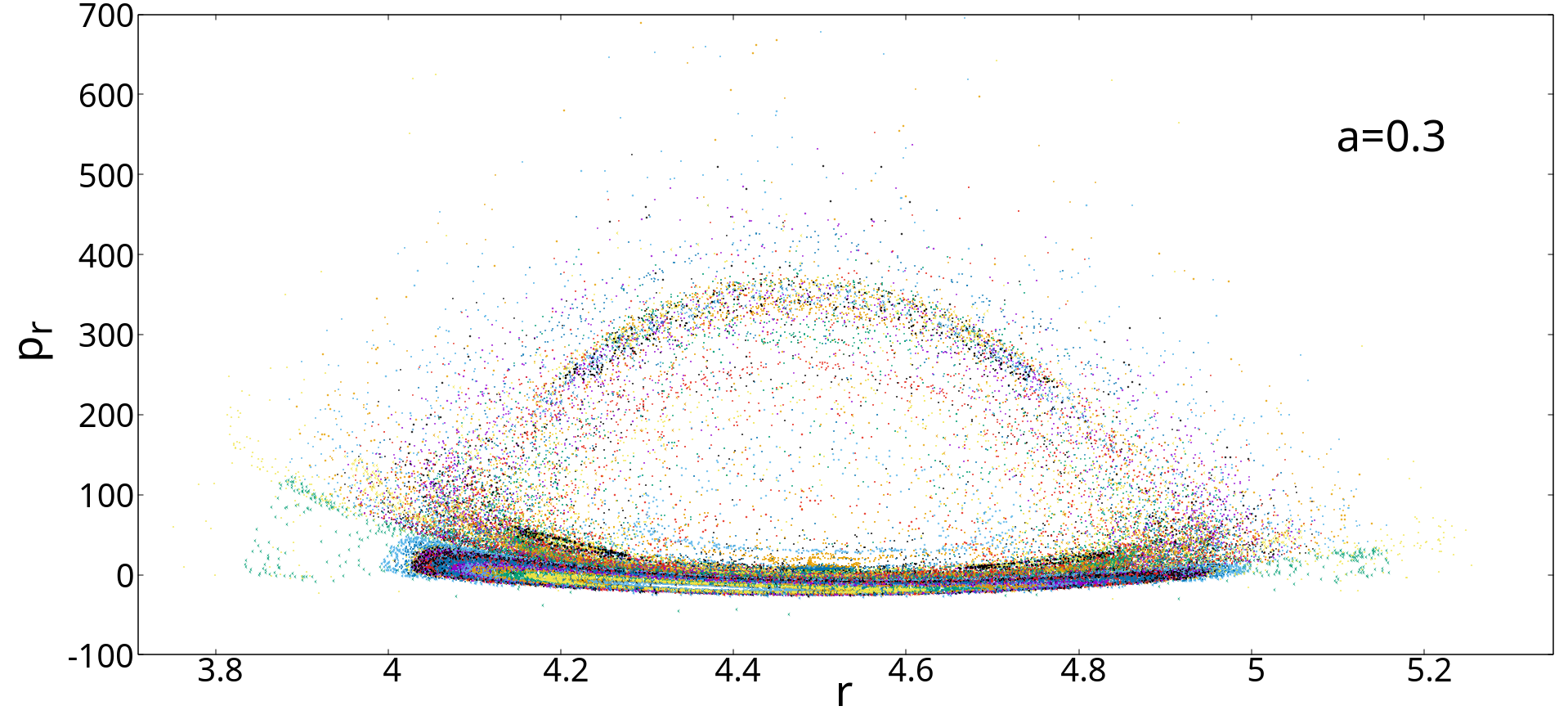}\label{6e}}
		\end{array}
		$\end{center}
	\caption{The Poincar$\Acute{e}$ sections in the $(r-p_r)$ plane with $p_{\theta}>0$ and $\theta=0$ with different dimensional parameter $a$ for a fixed energy $E=50$ for the SSS neutral black hole.}
	\label{f6}
\end{figure}

Furthermore, we analyze the nature of the trajectories by changing the dimensional parameter $a$ for a fixed energy $E=50$. In Fig.(\ref{f6}), we plot the Poincar$\Acute{e}$ sections in the plane $r-p_r$ for the different parameters $a=0.5,0.45,0.4,0.35$ and $0.3$ at fixed energy $E=50$. Here we have similar kinds of result i.e., for the higher values of $a$ (say $a=0.5,0.45$), there is a close region appears with the points in the phase plane. But as $a$ is started decreasing, the horizon interacts with the surface of the particles trajectories and as a consequences, the particles disappear and in parallel, the close region breaks due to the low value of the dimensional parameter $a$. Therefore, as a conclusion we clearly find that the decrease in dimensional parameter $a$ introduces chaotic fluctuations in the trajectories, and at very low value of the parameter (say $a=0.35,0.3$) the trajectory becomes fully chaotic in nature for the dynamic motions of the SSS charge-less black hole system. Thus the dimensional parameter $a$ has a key role due to the modification of gravity theory.



\section{Discussions and final remarks}\label{sec5}
Now, we summarize and conclude the final remarks of our work. In this study, we investigate the trajectories of a massless particle in the vicinity of a black hole horizon, focusing on the near-horizon region. Our theoretical analysis reveals that the radial motion of such a particle exhibits exponential growth over time, suggesting that the presence of the black hole horizon can induce chaotic behavior in an otherwise integrable system. Specifically, we explore the influence of a harmonic potential on the particle's motion and observe the emergence of chaos beyond a certain energy threshold. 

Static spherically symmetric spacetimes play a crucial role in the field of black hole physics due to their capacity to elucidate all fundamental attributes of black holes. This understanding can be extended to broader contexts, making them invaluable for generalization in various scenarios. The current study addresses two key facets of this topic. In the initial segment, we concentrate on the Lagrangian derivation of equations of motion for spherically symmetric spacetime within the framework of $f(R)$ gravitational theories. Consequently, we have obtained novel solutions for charged black hole as well as charge-less or neutral black hole solutions with remarkable properties in relation to specific forms of $f(R)=R-2a\sqrt{R}$, where $a$ is a dimensional parameter, signifies as a contribution in modified gravity. The principal advantages of these black holes lie in their reliance on the dimensional parameter $a$ and their dynamic Ricci scalar, that is found to be $R=\frac{1}{r^2}$. These novel solutions are distinct from the conventional general relativistic solutions, as the parameter $a$ is strictly prohibited from assuming a zero value. If one conduct a comprehensive analysis of the scalar invariance associated with these black holes, then he/she can finds that the Kretschmann scalar and Ricci tensor square invariants exhibit a dependence on the dimensional parameter $a$. Remarkably, all the invariants exhibit genuine singularities when the radial coordinate approaches zero $(r=0)$. This intriguing phenomenon underscores the unique characteristics and potential implications of these black hole solutions within the framework of $f(R)$ gravitational theory. It is also noteworthy that the observed distinction can be attributed to the unique nature of the charged and neutral black hole models obtained in our investigation. This black hole solutions arise within the framework of $f(R)$ gravity, a modified theory of gravity, and cannot be seamlessly reduced to General Relativity (GR). This departure from the standard gravitational theory underscores the need to consider alternative gravitational frameworks when examining the properties and behavior of black holes in diverse astrophysical settings. This research offers valuable insights into the potential ramifications of adopting non-standard gravity theories in the study of black holes.

In the next part, we extend our analysis to both SSS charged and neutral black holes, finding that chaos manifests within a specific energy range. Additionally, our findings highlight that the introduction of the dimensional parameter $a$ plays a crucial role to produce the chaos in the particle's motion. This research expands the understanding of the dynamics of particles near black hole horizons and underscores the general applicability of the SSS metric. Ultimately, our findings converge to an illuminating conclusion. While the dynamics of a particle in the presence of SSS charged and neutral black holes initially adhere to an integrable nature, the introduction of a harmonic perturbation leads to the emergence of chaos. This alignment with the Kolmogorov-Arnold-Moser (KAM) theory, which asserts that nonlinear perturbations in integrable systems breed chaos, resonates deeply. There are several examples reported earlier in support of this claim, such as the manifestation of chaos in the Henon–Heiles potential at high energy \cite{zotos} or the turbulent oscillations of a double pendulum under Hamiltonian influence \cite{okada,percival}. In our exploration, when the harmonic term remains minor, we discern the existence of regular tori. Yet, these tori progressively disintegrate into a scattered array of points as the system's energy escalates, indicative of the amplification in harmonic perturbation and the subsequent onset of chaos.

People have already shown that in the field of GR the horizon affects an integrable system to make it chaotic, however, in the diverting from the field of GR, in the case of modified theory of gravity, the effect of horizon on an integrable system still persists which is apparently visible from our work. Therefore, it suggests that whether in GR or in any modified version of gravity, if there is any presence of horizon, the system must try to be a chaotic one whenever it is under the grasp of the horizon. Chaos is inevitable in the region of horizon's influence. 
It's important to note that this is a suggestive rather than a definitive conclusion. Further investigations, potentially employing quantum mechanical methods, are warranted to delve deeper into this phenomenon and provide a more comprehensive understanding. Furthermore, it is of paramount significance to delve into the intriguing realm of chaotic particle dynamics influenced by the event horizon, extending our inquiry to encompass the diverse landscapes of modified gravity theories, including $f(R,T)$, $f(Q)$, $f(Q,T)$, $f(T)$, and various scalar-tensor theories of gravity. This multifaceted exploration promises to unveil captivating insights into the intricate interplay between gravity modifications and the emergence of chaotic motion near horizons, enriching our understanding of the fundamental dynamics in the cosmos.

\section*{Acknowledgment}
The authors gratefully acknowledge Prof. Subir Ghosh (Physics and Applied Mathematics Unit, Indian Statistical Institute, Kolkata) for various fruitful discussions, which utilized to represent this manuscript. S Das also would like to acknowledge all the members of PAMU, ISI, Kolkata for the hospitality during his long-term academic visit, where a part of this work has been done. S Dalui thanks the Department of Physics, Shanghai University for providing postdoctoral funds during the period of this work. Additionally, S Das thanks Raj Kumar Das (Senior Research Fellow, PAMU, ISI, Kolkata) for a helpful discussion.




\begin{thebibliography}{00}

\bibitem{Riess1}
A. G. Riess et al., Observational Evidence from Supernovae for an Accelerating Universe and a Cosmological Constant, \textcolor{blue}{Astron. J. 116, 1009 (1998)}.

\bibitem{Perlmutter1}
S. Perlmutter et al., Measurements of $\Omega$ and $\Lambda$ from 42 High-Redshift Supernovae, \textcolor{blue}{Astrophys. J. 517, 565 (1999)}.

\bibitem{De}
P. deBernardis et al., A flat Universe from high-resolution maps of the cosmic microwave background radiation, \textcolor{blue}{Nature 404, 955 (2000)}.

\bibitem{Colless}
M. Colless et al., The $2dF$ Galaxy Redshift Survey: spectra and redshifts, \textcolor{blue}{Mon. Not. R. Astron. Soc. 328, 1039 (2001)}.

\bibitem{Perlmutter2}
S. Perlmutter et al., New Constraints on $\Omega_{M}$, $\Omega_{\Lambda}$, and $\omega$ from an Independent Set of 11 High-Redshift Supernovae Observed with the Hubble Space Telescope, \textcolor{blue}{Astrophys. J. 598, 102 (2003)}.

\bibitem{Spergel1}
D. N. Spergel et al., First-Year Wilkinson Microwave Anisotropy Probe (WMAP) Observations: Determination of Cosmological Parameters, \textcolor{blue}{Astrophys. J. Suppl. Ser. 148, 175 (2003)}.

\bibitem{Peiris}
H. V. Peiris et al., First-Year Wilkinson Microwave Anisotropy Probe (WMAP) Observations: Implications For Inflation, \textcolor{blue}{Astrophys. J. Suppl. Ser. 148, 213 (2003)}.

\bibitem{Tegmark}
M. Tegmark et al., Cosmological parameters from SDSS and WMAP, \textcolor{blue}{Phys. Rev. D 69, 103501 (2004)}.

\bibitem{Cole}
S. Cole et al., The $2dF$ Galaxy Redshift Survey: power-spectrum analysis of the final data set and cosmological implications, \textcolor{blue}{Mon. Not. R. Astron. Soc. 362, 505 (2005)}.

\bibitem{Springel}
V. Springel et al., The large-scale structure of the Universe, \textcolor{blue}{Nature 440, 1137 (2006)}.

\bibitem{Astier}
P. Astier et al., The Supernova Legacy Survey: measurement of $\Omega_{\Lambda}$ and $\omega$ from the first year data set, \textcolor{blue}{Astron. Astrophys. 447, 31 (2006)}.

\bibitem{Riess2}
A. G. Riess et al., New Hubble Space Telescope Discoveries of Type Ia Supernovae at $z\geq 1$: Narrowing Constraints on the Early Behavior of Dark Energy*, \textcolor{blue}{Astrophys. J. 659, 98 (2007)}.

\bibitem{Spergel2}
D. N. Spergel et al., Three-Year Wilkinson Microwave Anisotropy Probe (WMAP) Observations: Implications for Cosmology, \textcolor{blue}{Astrophys. J. Suppl. Ser. 170, 377 (2007)}.

\bibitem{Ade}
P. A. R. Ade et al., Planck 2013 results. XVI. Cosmological parameters, \textcolor{blue}{Astron. Astrophys. 571, A16 (2014)}.

\bibitem{Komatsu}
E. Komatsu et al., Five-Year Wilkinson Microwave Anisotropy Probe (WMAP) Observations: Cosmological Interpretation, \textcolor{blue}{Astrophys. J. Suppl. Ser. 180 330 (2009)}.

\bibitem{capozziello} S. Capozziello and M. De Laurentis, Extended theories of gravity, \textcolor{blue}{Phys. Rep. 509, 167 (2011)}.

\bibitem{birrell} N. D. Birrell and P. C. W. Davies, Quantum Fields in Curved Space, \textcolor{blue}{Cambridge Monographs on Mathematical Physics (Cambridge University Press, Cambridge, England, 1984)}.

\bibitem{r1} S. Nojiri and S. D. Odintsov, Unified cosmic history in modified gravity: From $F(R)$ theory to Lorentz non-invariant models, \textcolor{blue}{Phys. Rep. 505, 59 (2011)}.

\bibitem{r2} S. Nojiri and S. D. Odintsov, Introduction to modified gravity and gravitational alternative for dark energy, \textcolor{blue}{Int. J. Geom. Methods Mod. Phys. 04, 115 (2007)}.

\bibitem{r3} S. Nojiri, S. D. Odintsov, and V. K. Oikonomou, Modified gravity theories on a nutshell: Inflation, bounce and late-time evolution, \textcolor{blue}{Phys. Rep. 692, 1 (2017)}.

\bibitem{r4} Thomas P. Sotiriou and V. Faraoni, $f(R)$ theories of gravity, \textcolor{blue}{Rev. Mod. Phys. 82, 451 (2010)}.

\bibitem{cai} Y. F. Cai, S. Capozziello, M. De Laurentis, and E. N. Saridakis, $f(T)$ teleparallel gravity and cosmology, \textcolor{blue}{Rep. Prog. Phys. 79, 106901 (2016)}.

\bibitem{c1} S. Capozziello, Curvature Quintessence, \textcolor{blue}{Int. J. Mod. Phys. D 11, 483 (2002)}.

\bibitem{c2} S. Nojiri and S. D. Odintsov, Modified gravity with negative and positive powers of curvature: Unification of inflation and cosmic acceleration, \textcolor{blue}{Phys. Rev. D 68, 123512 (2003)}.

\bibitem{c3} S. Nojiri and S. D. Odintsov, Where new gravitational physics comes from: M-theory?, \textcolor{blue}{Phys. Lett. B 576, 5 (2003)}.

\bibitem{c4} S. M. Carroll, V. Duvvuri, M. Trodden, and M. S. Turner, Is cosmic speed-up due to new gravitational physics?, \textcolor{blue}{Phys. Rev. D 70, 043528 (2004)}.

\bibitem{c5} G. Allemandi, A. Borowiec, and M. Francaviglia, Accelerated cosmological models in Ricci squared gravity, \textcolor{blue}{Phys. Rev. D 70, 103503 (2004)}.

\bibitem{c6} X. Meng and P. Wang, Cosmological evolution in $1/R$-gravity theory, \textcolor{blue}{Class. Quantum Gravit. 21, 951 (2004)}.

\bibitem{c7} S. M. Carroll, A. De Felice, V. Duvvuri, D. A. Easson, M. Trodden, and M. S. Turner, Cosmology of generalized modified gravity models, \textcolor{blue}{Phys. Rev. D 71, 063513 (2005)}.

\bibitem{c8} S. Capozziello and M. De Laurentis, Extended theories of gravity, \textcolor{blue}{Phys. Rep. 509, 167 (2011)}.

\bibitem{Ostro} M. Ostrogradski, \textcolor{blue}{Mem. Ac. St. Petersbourg VI 4, 385 (1850)}.

\bibitem{Wood} R. P. Woodard,  Avoiding dark energy with $1/R$ modifications of gravity, \textcolor{blue}{The Invisible Universe: Dark Matter and Dark Energy. Lecture Notes in Physics, 720, Springer, astro-ph/0601672 (2007)}.

\bibitem{Dol} A. D. Dolgov and M. Kawasaki, Can modified gravity explain accelerated cosmic expansion?, \textcolor{blue}{Phys. Lett. B 573, 1 (2003)}.

\bibitem{Sou} M. E. Soussa and R. P. Woodard, The force of gravity from a Lagrangian containing inverse powers of the Ricci scalar, \textcolor{blue}{Gen. Relativ. Gravit. 36, 855 (2004)}.

\bibitem{Faraoni} V. Faraoni, Stability of modified gravity models, \textcolor{blue}{Phys. Rev. D 72, 124005 (2005)}.

\bibitem{Chi} T. Chiba, $1/R$ gravity and scalar-tensor gravity, \textcolor{blue}{Phys. Lett. B 575, 1 (2003)}.

\bibitem{E} E. E. Flanagan, The conformal frame freedom in theories of gravitation, \textcolor{blue}{Class. Quantum Gravit. 21, 3817 (2004)}

\bibitem{Clif} T. Clifton and J. D. Barrow, The power of general relativity, \textcolor{blue}{Phys. Rev. D 72, 103005 (2005)}.

\bibitem{Mult1} T. Multamaki and I. Vilja, Cosmological expansion and the uniqueness of the gravitational action, \textcolor{blue}{Phys. Rev. D 73, 024018
(2006)}.

\bibitem{Dam} T. Damour and G. Esposito-Farese, Tensor-multi-scalar theories of gravitation, \textcolor{blue}{Class. Quantum Gravit. 9, 2093 (1992)}.

\bibitem{Mag} G. Magnano and L. M. Sokolowski, Physical equivalence between nonlinear gravity theories and a general-relativistic self-gravitating scalar field, \textcolor{blue}{Phys. Rev. D 50, 5039 (1994)}.


\bibitem{mult1} T. Multamäki and I. Vilja, Spherically symmetric solutions of modified field equations in $f(R)$ theories of gravity, \textcolor{blue}{Phys. Rev. D 74, 064022 (2006)}.

\bibitem{dela} A. de la Cruz-Dombriz, A. Dobado, A.L. Maroto, Black Holes in $f(R)$ theories,  \textcolor{blue}{Phys. Rev. D 80, 124011 (2009). (Erratum: [Phys. Rev. D 83, 029903 (2011)])}. 

\bibitem{moon} T. Moon, Y. S. Myung, E. J. Son, $f(R)$ black holes, \textcolor{blue}{Gen. Relativ. Gravit. 43, 3079 (2011)}.

\bibitem{jar} J. A. R. Cembranos, A. de la Cruz-Dombriz, P. Jimeno Romero, Kerr–Newman black holes in $f(R)$ theories, \textcolor{blue}{Int. J. Geom. Methods
Mod. Phys. 11, 1450001 (2014)}.

\bibitem{mult2} T. Multamäki and I. Vilja, Static spherically symmetric perfect fluid solutions in $f(R)$ theories of gravity, \textcolor{blue}{Phys. Rev. D 76, 064021 (2007)}.

\bibitem{sh1} S. H. Mazharimousavi, M. Halilsoy, T. Tahamtan, Constant curvature $f(R)$ gravity minimally coupled with Yang–Mills field, \textcolor{blue}{Eur. Phys. J. C 72, 1958 (2012)}.

\bibitem{sh2} S. H. Mazharimousavi, M. Halilsoy, Black hole solutions in $f(R)$
gravity coupled with non-linear Yang–Mills field, \textcolor{blue}{Phys. Rev. D 84, 064032 (2011)}.

\bibitem{habib} S. H. Mazharimousavi, M. Halilsoy, T. Tahamtan, Solutions for
$f(R)$ gravity coupled with electromagnetic field, \textcolor{blue}{Eur. Phys. J. C 72, 1851 (2012)}.

\bibitem{holl} L. Hollenstein, F.S.N. Lobo, Exact solutions of $f(R)$ gravity coupled to nonlinear electrodynamics, \textcolor{blue}{Phys. Rev. D 78, 124007 (2008)}.

\bibitem{rod} M. E. Rodrigues, E.L. Junior, G.T. Marques, V.T. Zanchin, Regular
black holes in $f(R)$ gravity coupled to nonlinear electrodynamics, \textcolor{blue}{Phys. Rev. D 94(2), 024062 (2016)}. 

\bibitem{hur} R. A. Hurtado, J. R. Arenas, Spherically symmetric and static solutions in $f(R)$ gravity coupled with EM fields, \textcolor{blue}{Phys. Rev. D 102, 104019, (2020)}.

\bibitem{sal} S. Capozziello, M. De laurentis, A. Stabile, Axially symmetric solutions in $f(R)$-gravity, \textcolor{blue}{Class. Quantum Gravity 27, 165008 (2010)}. 

\bibitem{hendi1} S. H. Hendi, The relation between $F(R)$ gravity and Einstein-conformally invariant Maxwell source, \textcolor{blue}{Phys. Lett. B 690, 220–223
(2010)}. 

\bibitem{hendi2} S. H. Hendi, B. E. Panah, S. M. Mousavi, Some exact solutions
of $F(R)$ gravity with charged (a)dS black hole interpretation, \textcolor{blue}{Gen. Relativ. Gravit. 44, 835 (2012)}. 

\bibitem{amirabi} Z. Amirabi, M. Halilsoy, S. H. Mazharimousavi, Generation of
spherically symmetric metrics in $f(R)$ gravity, \textcolor{blue}{Eur. Phys. J. C 76(6), 338 (2016)}.

\bibitem{trp} T. R. P. Carames, E. R. Bezerra de Mello, Spherically symmetric vacuum solutions of modified gravity theory in higher dimensions, \textcolor{blue}{Eur. Phys. J. C 64, 113–121 (2009)}.

\bibitem{main1} L. Sebastiani and S. Zerbini, Static spherically symmetric solutions in F(R) gravity, \textcolor{blue}{Eur. Phys. J. C 71, 1591 (2011)}.

\bibitem{main2} G. G. L. Nashed and S. Capozziello, Charged spherically symmetric black holes in $f(R)$ gravity and their stability analysis, \textcolor{blue}{Physical Review D 99, 104018 (2019)}.

\bibitem{saffari1} R. Saffari and S. Rahvar, $f(R)$ gravity: From the Pioneer anomaly to cosmic acceleration, \textcolor{blue}{Phys. Rev. D 77, 104028 (2008)}.

\bibitem{saffari2} R. Saffari and S. Rahvar, Consistency condition of spherically symmetric solutions in $f(R)$ gravity \textcolor{blue}{Mod. Phys. Lett. A 24, 305–309 (2009)}.

\bibitem{cognola} G. Cognola, M. Gastaldi, and S. Zerbini, On the stability of a class of modified gravitational models, \textcolor{blue}{Int. J. Theor. Phys. 47, 898 (2008)}

\bibitem{vilenkin} A. Vilenkin, Classical and quantum cosmology of the Starobinsky inflationary model \textcolor{blue}{Phys. Rev. D 32, 2511 (1985)}.

\bibitem{capo} S. Capozziello, Curvature Quintessence, \textcolor{blue}{Int. J. Mod. Phys. D 11, 4483 (2002)}.

\bibitem{abbott1} B. P. Abbott et al., Observation of Gravitational Waves from a Binary Black Hole Merger, \textcolor{blue}{Phys. Rev. Lett., 116, 061102 (2016)}.

\bibitem{abbott2} B. P. Abbott et al., GW151226: Observation of Gravitational Waves from a 22-Solar-Mass Binary Black Hole Coalescence, \textcolor{blue}{Phys. Rev. Lett., 116, 241103 (2016)}.

\bibitem{abbott3} B. P. Abbott et al., Binary Black Hole Mergers in the first Advanced LIGO Observing Run, \textcolor{blue}{Phys. Rev., 6, 041015 (2016)}.

\bibitem{abbott4} B. P. Abbott et al., GW170104: Observation of a 50-Solar-Mass Binary Black Hole Coalescence at Redshift 0.2, \textcolor{blue}{Phys. Rev. Lett., 118, 221101 (2017)}.

\bibitem{i1} L. Bombelli and E. Calzetta, Chaos around a black hole, \textcolor{blue}{Class. Quant. Grav. 9, 2573–2599 (1992)}.

\bibitem{i2} Y. Sota, S. Suzuki, and K.-i. Maeda, Chaos in static axisymmetric spacetimes. 1:Vacuum case, \textcolor{blue}{Class. Quant. Grav. 13, 1241–1260 (1996)}.

\bibitem{i3} W. M. Vieira and P. S. Letelier, Chaos around a Henon-Heiles inspired exact perturbation of a black hole, \textcolor{blue}{Phys. Rev. Lett. 76, 1409–1412 (1996)}.

\bibitem{i4} S. Suzuki and K.-i. Maeda, Chaos in Schwarzschild spacetime: The motion of a spinning particle, \textcolor{blue}{Phys. Rev. D 55, 4848–4859 (1997)}.

\bibitem{i5} N. J. Cornish and N. E. Frankel, The Black hole and the pea, \textcolor{blue}{Phys. Rev. D 56, 1903–1907 (1997)}.

\bibitem{i6} A. P. S. de Moura and P. S. Letelier, Chaos and fractals in geodesic motions around a non-rotating black hole with an external halo, \textcolor{blue}{Phys. Rev. E 61, 6506–6516 (2000)}.

\bibitem{i7} M. D. Hartl, Dynamics of spinning test particles in Kerr spacetime, \textcolor{blue}{Phys. Rev. D 67, 024005 (2003)}.

\bibitem{i8} W. Han, Chaos and dynamics of spinning particles in Kerr spacetime, \textcolor{blue}{Gen. Rel. Grav. 40, 1831–1847, (2008)}.

\bibitem{i9} M. Takahashi and H. Koyama, Chaotic motion of Charged Particles in an Electromagnetic Field Surrounding a Rotating Black Hole, \textcolor{blue}{Astrophys. J. 693, 472–485 (2009)}.

\bibitem{i10} D. Li and X. Wu, Chaotic motion of neutral and charged particles in a magnetized Ernst-Schwarzschild spacetime, \textcolor{blue}{Eur. Phys. J. Plus 134, 96 (2019)}.

\bibitem{adhikari} A. Adhikari, K. Bhattacharya, C. Chowdhury, B. R. Majhi, Fluctuation–dissipation relation in accelerated frames, \textcolor{blue}{Phys. Rev. D 97, 045003 (2018)}.

\bibitem{hashimoto} K. Hashimoto and N. Tanahashi, Universality in Chaos of Particle Motion near Black Hole Horizon, \textcolor{blue}{Phys. Rev. D 95, 024007 (2017)}.

\bibitem{parikh} M. Parikh, F. Wilczek, Hawking radiation as tunneling, \textcolor{blue}{Phys. Rev. Lett. 85, 5042 (2000)}.

\bibitem{rabin} R. Banerjee, B. R. Majhi, Quantum tunneling and back reaction, \textcolor{blue}{Phys. Lett. B 662, 62 (2008)}.

\bibitem{zhang} Y. Zhang, X. Liu, M. R. Belic, W. Zhong, Y. Zhang, M. Xiao, Propagation dynamics of a light beam in a fractional Schrödinger equation, \textcolor{blue}{Phys. Rev. Lett. 115, 180403 (2015)}.

\bibitem{kow} K. Kowalski, J. Rembielinski, The relativistic massless harmonic oscillator, \textcolor{blue}{Phys. Rev. A 81, 012118 (2010)}.

\bibitem{safko} J. L. Safko, F. Elston, Lagrange multipliers and gravitational theory, \textcolor{blue}{J. Math. Phys. 17, 1531–1537 (1976)}.

\bibitem{capo1} S. Capozziello, A. Stabile, A. Troisi, Spherical symmetry in $f(R)$-gravity, \textcolor{blue}{Class. Quantum Gravity 25, 085004 (2008)}.

\bibitem{blau} M. Blau, Lecture notes on General Relativity, \textcolor{blue}{http://www.blau.itp.unibe.ch/GRLecturenotes.html, 2023}.


\bibitem{carroll} S. M. Carroll, An Introduction to General Relativity: Spacetime and Geometry, \textcolor{blue}{Addison Wesley, New York, 2004}.

\bibitem{maldacena} J. Maldacena, S. H. Shenker, D. Stanford, A bound on chaos, \textcolor{blue}{J. High Energy Phys. 1608, 106 (2016)}.

\bibitem{dalui} S. Dalui, B. R. Majhi, and P. Mishra, Presence of horizon makes particle motion chaotic, \textcolor{blue}{Phys. Lett. B 788 486–493 (2019)}.

\bibitem{Dalui:2019umw}
S. Dalui, B. R. Majhi and P. Mishra, Induction of chaotic fluctuations in particle dynamics in a uniformly accelerated frame, \textcolor{blue}{Int. J. Mod. Phys. A \textbf{35} (2020) no.18, 2050081}. 

\bibitem{KAM} J. Guckenheimer, P. Holmes, Nonlinear Oscillations, Dynamical Systems and Bifurcations of Vector Fields, \textcolor{blue}{Springer-Verlag, New York, NY, (2002)}.

\bibitem{zotos} E. E. Zotos, Classifying orbits in the classical Henon–Heiles Hamiltonian system, \textcolor{blue}{Nonlinear Dyn. (NODY) 79 1665–1677 (2015)}.

\bibitem{okada} T. Stachowiak, T. Okada, A numerical analysis of chaos in the double pendulum, \textcolor{blue}{Chaos Solitons Fractals 29 (2) 417–422 (2006)}.

\bibitem{percival} I. C. Percival, Chaos in Hamiltonian systems, \textcolor{blue}{Proc. R. Soc. Lond. Ser. A, Math. Phys. Sci. 413, 131–143 (1987)}.


\end{thebibliography}
\end{document}